\documentclass[aps,prd,twocolumn,showpacs,superscriptaddress,nofootinbib,preprintnumbers]{revtex4-2}

\usepackage{graphicx}
\usepackage{color}
\usepackage{amsmath}
\usepackage{amssymb}
\usepackage{enumerate}
\usepackage[colorlinks=true,citecolor=blue,urlcolor=blue]{hyperref}

\graphicspath{ {Figures/} }

\newcommand{\efac}[1]{S_{#1}}
\newcommand{\equad}[1]{Q_{#1}}

\begin{document}

\title{Harmonic analysis for pulsar timing arrays}

\begin{abstract}
\noindent We investigate the use of harmonic analysis techniques to perform measurements of the angular power spectrum on mock pulsar timing data for an isotropic stochastic gravitational-wave background (SGWB) with a dimensionless strain amplitude $A_{\text{gw}}=2 \times 10^{-15}$ and spectral index $\gamma_{\text{gw}}=13/3$. We examine the sensitivity of our harmonic analysis to the number of pulsars (50, 100, and 150) and length of pulsar observation time (10, 20, and 30 years) for an isotropic distribution of pulsars. We account for intrinsic pulsar red noise and use an average value of white noise of $\sim 100\:{\rm ns}$. We are able to detect the quadrupole for all our mock harmonic analyses, and for 150 pulsars observed for 30 years we are able to detect up to the $\ell = 5$ multipole. We provide linear scaling relationships for the SGWB amplitude, the quadrupole, and $\ell = 3$ as a function of pulsar observation time and number of pulsars. We estimate the sensitivity of our harmonic approach to deviations of general relativity that produce subluminal GW propagation speeds.
\end{abstract}

\author{Jonathan Nay}
\affiliation{Texas Center for Cosmology and Astroparticle Physics, Weinberg Institute, Department of Physics, The University of Texas at Austin, Austin, TX 78712, USA}
\author{Kimberly K.~Boddy}
\affiliation{Texas Center for Cosmology and Astroparticle Physics, Weinberg Institute, Department of Physics, The University of Texas at Austin, Austin, TX 78712, USA}
\author{Tristan L.~Smith}
\affiliation{Department of Physics and Astronomy, Swarthmore College, Swarthmore, PA 19081, USA}
\author{Chiara M.~F.~Mingarelli}
\affiliation{Department of Physics, University of Connecticut, 196 Auditorium Road, U-3046, Storrs, CT, 06269-3046, USA}
\affiliation{Center for Computational Astrophysics, Flatiron Institute, 162 Fifth Ave, New York, NY, 10010, USA}
\affiliation{Department of Physics, Yale University, New Haven, CT, 06520, USA}

\maketitle

%%%%%%%%%%%%%%%%%%%%%%%%%%%%%%%%%%%%%%%%%%%%%%%%%%%%%%%%%%%%%%%%%%%%%%%%%%%%%%%
\section{Introduction}
\label{sec:intro}

The direct detection of gravitational waves (GWs) by LIGO~\cite{LIGOScientific:2016aoc} has marked the rise of an exciting new observational era.
Currently, all reported GW measurements are of individual binary mergers of compact objects, but there are also ongoing efforts to detect a stochastic GW background (SGWB) arising from a population of unresolved binaries.%
\footnote{There may also be contributions from a cosmological SGWB, sourced by exotic processes in the early Universe~\cite{Boddy:2022knd, Caldwell:2022qsj, Green:2022hhj}.}
In particular, pulsar timing arrays (PTAs)~\cite{Sazhin:1978, Detweiler:1979wn, Maggiore:2018sht, MingCC2022} are searching for a SGWB in the nanohertz regime, generated by mergers of supermassive black hole binaries.
PTAs measure pulses of radio emission from millisecond pulsars, which serve as precise astronomical clocks due to their highly stable rotational periods.
Intervening GWs between Earth and a pulsar (typically $\sim$ a kiloparsec away) induce small shifts in the pulse times of arrival (TOAs), and PTA experiments achieve sensitivity to the effects of GWs by cross-correlating TOA information between pairs of pulsars.

An isotropic SGWB imprints two main signatures in TOA data: low-frequency timing shifts common to \emph{all} observed pulsars (referred to as ``common red noise'' or a ``common process'') and an angular correlation of timing shifts between pulsar pairs, known as the Hellings--Downs (HD) curve~\cite{Hellings:1983fr}.
Thus far, multiple PTA collaborations have reported an observed common red noise, broadly consistent with expectations from a SGWB, but have not found evidence of the HD correlation necessary to claim a detection~\cite{Antoniadis:2022pcn, NANOGrav:2020bcs, Goncharov:2021oub, Chalumeau:2021fpz}.%
\footnote{However, see the note added.}

Standard PTA pipelines incorporate a Bayesian analysis to characterize frequency spectrum information and a frequentist analysis~\cite{Anholm:2008wy} to assess the evidence of HD angular correlations (e.g., see Refs.~\cite{Antoniadis:2022pcn, NANOGrav:2020bcs, Goncharov:2021oub, Chalumeau:2021fpz}).
However, an equivalent way to represent the HD angular correlation function is through the angular power spectrum, obtained directly by decomposing the isotropic SGWB into spherical harmonics~\cite{Dai:2012bc} with multipole $\ell$~\cite{Qin:2018yhy}.
The angular power spectrum for an isotropic SGWB has a dominant quadrupole ($\ell=2$) contribution due to the tensorial nature of GWs, while higher multipole contributions scale as $\sim$$\ell^{-4}$~\cite{ Gair:2014rwa, Qin:2018yhy}.

In this paper, we investigate the capabilities of PTAs to measure the angular power spectrum of an isotropic SGWB.
We perform Bayesian analyses with mock PTA data and allow the relative amplitudes of the angular power spectrum multipoles to vary as independent parameters.
This flexibility naturally permits a search of an isotropic SGWB that incorporates generic modifications of general relativity (GR), which may feature GWs with subliminal speeds and non-Einsteinian polarization modes~\cite{Mihaylov:2018uqm, Qin:2018yhy, Mihaylov:2019lft, Qin:2020hfy, Bernardo:2022rif}.
By contrast, standard PTA analyses assume HD correlations, which correspond to an angular power spectrum with specific fixed values for the multipole amplitudes.
Measuring the shape of the angular power spectrum may also help readily identify the presence of other correlations, such as monopolar clock errors and dipolar solar system ephemeris errors~\cite{NANOGrav:2020bcs}, as well as anisotropies in the SGWB~\cite{Mingarelli:2013dsa, Taylor:2013esa, Gair:2014rwa, Hotinli:2019tpc}.

In order to assess the ability for PTAs to extract the angular power spectrum, we analyze various mock realizations of PTA data assuming standard GR and an isotropic SGWB with a dimensionless strain amplitude $A_{\text{gw}}=2 \times 10^{-15}$ and spectral index $\gamma_{\text{gw}}=13/3$.
We consider different numbers of observed pulsars (50, 100, and 150) and lengths of observation time (10, 20, and 30 years) for an isotropic distribution of pulsars.
In addition to the overall SGWB amplitude, we treat the relative amplitudes for each multipole from $\ell =2$ to $\ell =8$ as independent parameters.
We also include intrinsic red and white noise for each pulsar and assume the noise is subdominant (in the low-frequency regime of interest) to the SGWB signal, with an average white-noise value of $\sim 100 \: {\rm ns}$.

In all our harmonic analyses, we are able to detect the quadrupole.
For our harmonic analysis with 150 pulsars observed for 30 years, we are able to detect multipoles up to $\ell = 5$.
The first four nonzero multipoles in the angular power spectrum can accurately reconstruct the HD curve due to the sharp dropoff in angular power strength as $\ell$ increases, so our harmonic analysis approach is a promising tool to help characterize angular correlations present in pulsar timing data.

For the SGWB amplitude, the quadrupole, and $\ell = 3$, we provide linear scaling relationships as a function of experiment observation time and as a function of the number of pulsars in our model.
We find that increasing the observation time has a larger scaling effect than increasing the number of pulsars.
Longer observation times correspond to accessing lower-frequency GWs, where the GW strain is larger (e.g., for a SGWB arising from supermassive black hole binaries~\cite{Phinney:2001di}).

Finally, we give an example of how our harmonic analysis method enables us to explore deviations from GR by considering subluminal GW propagation speeds.
For our harmonic analysis with 150 pulsars observed for 30 years, we find the constraints vary by multipole and range from $0.95c$ to $0.98c$, where $c$ is the speed of light.

This paper is organized as follows.
In Sec.~\ref{sec:background} we review how PTA timing data are modeled and define the angular power spectrum.
In Sec.~\ref{sec:analysis-harmonic} we present our methods for performing a Bayesian harmonic analysis.
We describe how we generate our mock PTA data and show our results in Sec.~\ref{sec:analysis}.
We conclude in Sec.~\ref{sec:conclusions}.
We also include various supplementary material.
In Appendix~\ref{app:seeds} we demonstrate our results are not driven by numerical outliers when generating mock data by performing a realization study in which we vary the pseudorandom number generator seeds.
In Appendix~\ref{app:savage-dickey} we show our method for calculating a Savage-Dickey Bayes factor as a measure of the evidence for each multipole in our model.
In Appendix~\ref{app:corner} we provide corner plots showing the marginalized 1D and 2D posterior distributions for all SGWB model parameters of the harmonic analyses presented in Sec.~\ref{sec:analysis}.

%%%%%%%%%%%%%%%%%%%%%%%%%%%%%%%%%%%%%%%%%%%%%%%%%%%%%%%%%%%%%%%%%%%%%%%%%%%%%%%
\section{Pulsar timing residuals}
\label{sec:background}

PTAs time millisecond pulsars by recording pulses over a window of time of $\lesssim$ few days.
Within this window, TOAs are obtained (at multiple radio telescope frequencies) by integrating the pulses over $\sim$1 hour in order to increase the signal-to-noise ratio.
However, for the purposes of this paper, we consider a wideband approach~\cite{NANOGrav:2020qll}, for which there is a single TOA associated with the full window of time, and we use a single radio telescope frequency, since we are not considering the effects of the dispersion measure on the timing signal.
A pulsar timing residual is then obtained by fitting out all known systematic and deterministic processes from the TOA (e.g., see Ref.~\cite{NANOGrav:2020gpb}).
PTAs accumulate many timing residuals over the lifetime of the experiment ($\sim$10 years) by observing a given pulsar with a cadence of 2--3 weeks.

The $i^{th}$ timing residual observed at time $t_{ai}$ from pulsar $a$ located in the $\hat{n}_a$ direction can be expressed as (e.g., see Eq.~23.59 of Ref.~\cite{Maggiore:2018sht})
\begin{equation}
  r_a(\hat{n}_a,t_{ai}) = n_a(t_{ai}) + R^{\text{gw}}(\hat{n}_a,t_{ai}) + R_a^{\text{RN}}(t_{ai}) + R^{\text{det}}_{ai},
  \label{eq:TimingResidual}
\end{equation}
where $n_a(t_{ai})$ is Gaussian white noise, $R^{\text{gw}} (\hat{n}_a,t_{ai})$ is the contribution from a SGWB, $R_a^{\text{RN}}(t_{ai})$ is the contribution from pulsar intrinsic red noise, and $R^{\text{det}}_{ai}$ is any leftover deterministic signal not properly fit out from the TOA~\cite{Blandford:1984, Lentati:2013rla}.
The contribution from $R^{\text{det}}_{ai}$ is accounted for in our analyses but does not otherwise play an important role for our investigations, so we do not discuss it in detail.

In general, $R^{\text{gw}}(\hat{n}_a,t_{ai})$ also depends on the distance from Earth to the pulsar.
However, as we discuss below, we are interested in the two-point correlation function of $R^{\text{gw}}(\hat{n}_a,t_{ai})$ between pairs of pulsars, and the terms involving pulsar distances essentially only contribute for colocated pulsars, doubling the correlation function at vanishing separation (see Sec. 23.3 of Ref.~\cite{Maggiore:2018sht} for further discussion).
The doubling is commonly referred to as the ``pulsar term,'' which does not depend on the pulsar distances.
This approach is consistent with the methodology used by PTA collaborations when searching for a SGWB (e.g., see Ref.~\cite{NANOGrav:2020bcs}) and is justified for actual pulsar distances observed by PTAs~\cite{Mingarelli:2014xfa}.

For the remainder of this section, we discuss our modeling of the Gaussian white noise, the red-spectrum processes stemming from a SGWB and pulsar red noise, and the angular power spectrum of the SGWB, which is the foundation of our harmonic analysis approach.

%%%%%%%%%%%%%%%%%%%%%%%%%%%%%%%%%%%%%%%%
\subsection{Gaussian white noise}

The Gaussian white noise $n_a(t_{ai})$ is taken to have zero mean with a covariance matrix
\begin{equation}
  N_{ai, bj} \equiv \left< n_a(t_{ai}) \: n_b(t_{bj}) \right> = \sigma^{2}_{\text{WN},ai} \: \delta_{ab} \: \delta_{ij},
  \label{eq:WNCovariance}
\end{equation}
where $\delta_{ab}$ and $\delta_{ij}$ are Kronecker delta functions, and the angle brackets denote an ensemble average.
The total white-noise variance of the $i^{th}$ TOA measurement for pulsar $a$ is (e.g., see Sec. 3.3 of Ref.~\cite{NANOGrav:2020gpb}):
\begin{equation}
  \sigma^{2}_{\text{WN},ai} = \efac{a}^2 \left( \sigma^2_{\text{TOA},ai} + \equad{a}^2 \right),
  \label{eq:WNvariance}
\end{equation}
where $\sigma_{\text{TOA},ai}$ is the TOA measurement error, $\efac{a}$ is the instrument scaling error (also referred to as EFAC), and $\equad{a}$ is the instrument quadrature error (EQUAD).
We assume a single instrument measures all TOAs for a given pulsar, so there is only one scaling and one quadrature error per pulsar.

Since we assume a wideband approach, in which a single TOA is obtained over a single observation window of a pulsar, the white noise of different TOA measurements should not be correlated (e.g., there is no pulse phase ``jitter'').
The lack of such a correlated error (ECORR) is represented by the diagonal nature of Eq.~\eqref{eq:WNCovariance}.

%%%%%%%%%%%%%%%%%%%%%%%%%%%%%%%%%%%%%%%%
\subsection{Red-spectrum processes}
\label{sec:red-processes}

The red-spectrum processes due to a SGWB and pulsar intrinsic red noise are assumed to be stationary and Gaussian (e.g., see Ref.~\cite{vanHaasteren:2014qva}) with zero mean.
As with the Gaussian white noise, these processes are thus fully characterized by the two-point correlation functions
\begin{align}
  \left\langle R^{\text{gw}} (\hat{n}_a,t_{ai}) \: R^{\text{gw}} (\hat{n}_b,t_{bj})\right\rangle &= \xi^{\text{gw}}(\tau) \: \zeta(\cos{\theta_{ab}}) \label{eq:GWCovariance} \\
  \left\langle R_a^{\text{RN}}(t_{ai}) \: R_b^{\text{RN}}(t_{bj}) \right\rangle &= \xi^{\text{RN}}_a (\tau) \: \delta_{ab}, \label{eq:RNCovariance}
\end{align}
where $\tau \equiv |t_{ai}- t_{bj}|$, $\cos{\theta_{ab}} = \hat{n}_a \cdot \hat{n}_b$ is the pulsar-pair separation angle, and we have assumed the time dependence and angular dependence of the SGWB correlation function are separable [e.g., see Eqs.~(23.60) and (23.64) of Ref.~\cite{Maggiore:2018sht}].
Also, since we do not model distances to the pulsars, the condition $\theta_{ab}=0$ is equivalent to $a=b$.
For an isotropic SGWB, the spatial correlation function $\zeta(\cos{\theta_{ab}})$ is the standard HD curve~\cite{Hellings:1983fr}
\begin{align}
  \zeta(\cos{\theta_{ab}}) & = \frac{1}{2} (1 + \delta_{ab}) - \frac{1}{4} \left(\frac{1-\cos{\theta_{ab}}}{2} \right) \nonumber \\
  & + \frac{3}{2} \left(\frac{1-\cos{\theta_{ab}}}{2} \right) \: \log{\left(\frac{1-\cos{\theta_{ab}}}{2} \right)},
  \label{eq:HDcurve}
\end{align}
which has been normalized to 1 for $a=b$. The factor $\delta_{ab}$ accounts for an enhancement in the autocorrelation due to the pulsar term [e.g., see Eq.~(23.65) of Ref.~\cite{Maggiore:2018sht}].

From the Wiener-Khinchin theorem, $\xi^{\text{gw}}$ and $\xi^{\text{RN}}_a$ each possess a spectral decomposition given by a frequency power spectrum.
We parameterize the frequency power spectra for the SGWB and the intrinsic pulsar red noise as power laws of the form~\cite{Antoniadis:2022pcn, NANOGrav:2020bcs, Goncharov:2021oub, Chalumeau:2021fpz}
\begin{align}
  P^{\text{gw}}(f) &\equiv \frac{A^{2}_{\text{gw}}}{12 \pi^{2}} \: \left(\frac{f}{f_{\text{yr}}}\right)^{-\gamma_{\text{gw}}}\: f_{\text{yr}}^{-3},  \label{eq:GWpowerlaw} \\
  P^{\text{RN},a}(f) &\equiv \frac{A^{2}_{\text{RN},a}}{12 \pi^{2}} \: \left(\frac{f}{f_{\text{yr}}}\right)^{-\gamma_{\text{RN},a}}\: f_{\text{yr}}^{-3},
  \label{eq:RNpowerlaw}
\end{align}
where $A_{\text{gw}}$ is the dimensionless strain amplitude of the SGWB at a reference frequency $f_{\text{yr}}=1/\text{yr}$, $A_{\text{RN},a}$ is an equivalent dimensionless amplitude for the intrinsic red noise of pulsar $a$, and $\gamma_{\text{gw}}$ and $\gamma_{\text{RN},a}$ are the corresponding spectral indices.
For a source of inspiraling supermassive black hole binaries, we expect $\gamma_{\text{gw}} \simeq 13/3$~\cite{Phinney:2001di}.
We assume $\gamma_{\text{RN},a}$ and $\gamma_{\text{gw}}$ are positive, resulting in red spectra, which have more power at lower frequencies.
The SGWB spectrum in Eq.~\eqref{eq:GWpowerlaw} is common to all pulsars and thus does not carry the pulsar label $a$.

We use the same method of modeling the autocorrelation of red-spectrum processes as are used by PTA collaborations to search for a SGWB (e.g., see Ref.~\cite{NANOGrav:2020bcs}).
The difference with our harmonic analysis approach is how we model the cross-correlations (i.e., the correlations created from distinct pulsar pairs), which we discuss in the following section.

%%%%%%%%%%%%%%%%%%%%%%%%%%%%%%%%%%%%%%%%
\subsection{Angular power spectrum}

The spatial correlation function $\zeta(\cos\theta_{ab})$ for distinct pulsars $a$ and $b$ can be written as a Legendre polynomial decomposition~\cite{Burke:1975zz, Mingarelli:2013dsa, Gair:2014rwa, Roebber:2016jzl}
\begin{align}
  \zeta(\cos\theta_{ab})  = \sum_{\ell = 2}^{\infty} c_\ell P_\ell(\cos\theta_{ab}),
  \label{eq:LegendreCoefficient}
\end{align}
where $P_{\ell}(\cos\theta_{ab})$ are Legendre polynomials and $c_\ell$ are the associated Legendre coefficients.
The first nonzero Legendre coefficient is the quadrupole $(\ell=2)$ term.
The Legendre coefficients that reconstruct the HD curve for an isotropic SGWB are
\begin{equation}
  c_{\ell} = \frac{3}{2} \: (2\ell+1) \: \frac{(\ell-2)!}{(\ell+2)!},
  \label{eq:LegendreCoefficientscale}
\end{equation}
which exhibit a dominant quadrupolar contribution and a sharp reduction at higher multipoles.
The Legendre coefficients in Eq.~\eqref{eq:LegendreCoefficientscale} assume standard GR.
Modifications of GR require generalized Legendre coefficients, as shown in Refs.~\cite{Mihaylov:2018uqm,Qin:2018yhy,Mihaylov:2019lft,Qin:2020hfy,Bernardo:2022rif}.
By using a total-angular-momentum basis~\cite{Dai:2012bc} for the GW metric perturbation, the generalized Legendre coefficients can be written as~\cite{Qin:2018yhy,Qin:2020hfy}
\begin{equation}
  c_\ell \equiv 3(2\ell + 1) |
  F_\ell|^2,
  \label{eq:LegendreCoefficientGeneral}
\end{equation}
where $|F_\ell|^2$ is the detector response function defined by Eq.~(19) of Ref.~\cite{Qin:2018yhy}.
From Eqs.~\eqref{eq:LegendreCoefficientscale} and~\eqref{eq:LegendreCoefficientGeneral}, the detector response function under standard GR is
\begin{equation}
  |F_\ell|^2 = \frac{1}{2} \frac{(\ell - 2)!}{(\ell + 2)!}.
  \label{eq:GRresponse}
\end{equation}
For an isotropic SGWB, the detector response function is related to the angular power spectrum, $C_\ell(\tau)$, by~\cite{Qin:2018yhy,Qin:2020hfy}%
\footnote{In practice, the frequency integral in Eq.~\eqref{eq:SGWBAngularPS} employs a lower and upper frequency cutoff, which we provide in Sec.~\ref{sec:analysis-harmonic}.}
\begin{equation}
  C_{\ell}(\tau) = 12\pi |F_\ell|^2
  \times 2\int_{0}^\infty df\: P^{\text{gw}}(f)\: \cos(2\pi f \tau).
  \label{eq:SGWBAngularPS}
\end{equation}

%%%%%%%%%%%%%%%%%%%%%%%%%%%%%%%%%%%%%%%%%%%%%%%%%%%%%%%%%%%%%%%%%%%%%%%%%%%%%%%
\section{Harmonic Analysis}
\label{sec:analysis-harmonic}

In this section, we present our analysis method for measuring the angular power spectrum of an isotropic SGWB.
Standard PTA analyses fix the angular correlations of timing residuals to follow the HD curve in Eq.~\eqref{eq:HDcurve}.
We extend these analysis pipelines by replacing the assumed HD correlation function with the more general form given in Eq.~\eqref{eq:LegendreCoefficient} and treating the Legendre coefficients $c_\ell$ as independent parameters.
This generalized approach permits a measurement of the angular power spectrum directly from data, allowing for a more generic search of a SGWB that is agnostic to, e.g., possible modifications of GR.

We incorporate a finite number of multipoles as new parameters, with the general expectation that contributions from higher multipoles are suppressed and can be neglected.
As we describe in Sec.~\ref{sec:mock}, we generate mock data assuming an isotropic SGWB under standard GR, and we find that including coefficients for multipoles (starting at $\ell=2$) up to $\ell=8$ is sufficient in our analyses.
Note that we could also include the coefficients for $\ell=0$ and $\ell=1$, which may arise from clock or solar system ephemeris errors~\cite{NANOGrav:2020bcs}, respectively, or from non-Einsteinian polarization modes.
Monopole and dipole contributions from non-Einsteinian modes possess their own amplitudes that generally differ from the quadrupolar contribution from GR, while clock and ephemeris errors are expected to have completely different frequency spectra.
We leave explorations of monopole and dipole correlations to future work.

For our Bayesian analysis, we consider the likelihood function (e.g., see Ref.~\cite{NANOGrav:2015aud})
\begin{equation}
  p(\vec{r} \: | \: \vec{\eta}) = \frac{1}{\sqrt{\text{det}(2 \pi \mathcal{C})}} e^{-\frac{1}{2}\vec{r}^T \mathcal{C}^{-1} \vec{r}},
  \label{eq:Likelihood}
\end{equation}
where $\vec{r}$ is a vector of measured timing residuals with entries represented by Eq.~\eqref{eq:TimingResidual}, $\vec{\eta}$ represents the model parameters of interest, and $\mathcal{C}$ is an analytically marginalized covariance matrix.
$\mathcal{C}$ includes white-noise and red-noise covariances in Eqs.~\eqref{eq:WNCovariance} and \eqref{eq:RNCovariance}, as well as the covariance in Eq.~\eqref{eq:GWCovariance} from cross-correlations induced by GWs.

As we describe in Sec.~\ref{sec:background}, the $\tau$-dependent covariance functions in Eqs.~\eqref{eq:GWCovariance} and \eqref{eq:RNCovariance} are related to their corresponding frequency power spectra via the Wiener-Khinchin theorem [cf.\ Eq.~\eqref{eq:GWpowerlaw}].
However, in practice, PTA analyses implement the spectral decomposition using a finite number of sine and cosine basis functions consisting of the lowest harmonics of the fundamental frequency $f_L = 1/T_{{\rm obs}}$, where $T_{{\rm obs}}$ is the total observation time of the experiment~\cite{vanHaasteren:2014qva}.
Using the same methodology as PTA collaborations (e.g., see Refs.~\cite{NANOGrav:2020bcs, Antoniadis:2022pcn}), we use a high-frequency cutoff of $f_{H, {\rm gw}} = 14/15~\text{yr}^{-1}$ for the SGWB and $f_{H, {\rm RN}} = 30/15~\text{yr}^{-1} = 2~\text{yr}^{-1}$ for the intrinsic red noise.

%%%%%%%%%%%%%%%%%%%%%%%%%%%%%%%%%%%%%%%%%%%%%%%%%%%%%%%%%%%%%%%%%%%%%%%%%%%%%%%
\section{Methods and results}
\label{sec:analysis}

In order to assess the performance of a harmonic analysis, we want to understand how well the angular power spectrum can be reconstructed under various observational scenarios.
In this section, we describe the generation of mock data, outline our analysis pipeline, and present our results.

%%%%%%%%%%%%%%%%%%%%%%%%%%%%%%%%%%%%%%%%%%%%%%%%%%%%%%%%%%%%%%%%%%%%%%%%%%%%%%%
\subsection{Generating mock data}
\label{sec:mock}

We create nine different mock PTA datasets with varying observation time $T_{{\rm obs}}$ (10, 20, and 30 years) and number of pulsars $N_p$ (50, 100, and 150).
Our mock data are generated using a method which is similar to that used in Ref.~\cite{NANOGrav:2020spf}.

\begin{figure}[t]
  \centering
  \includegraphics[width=0.48\textwidth]{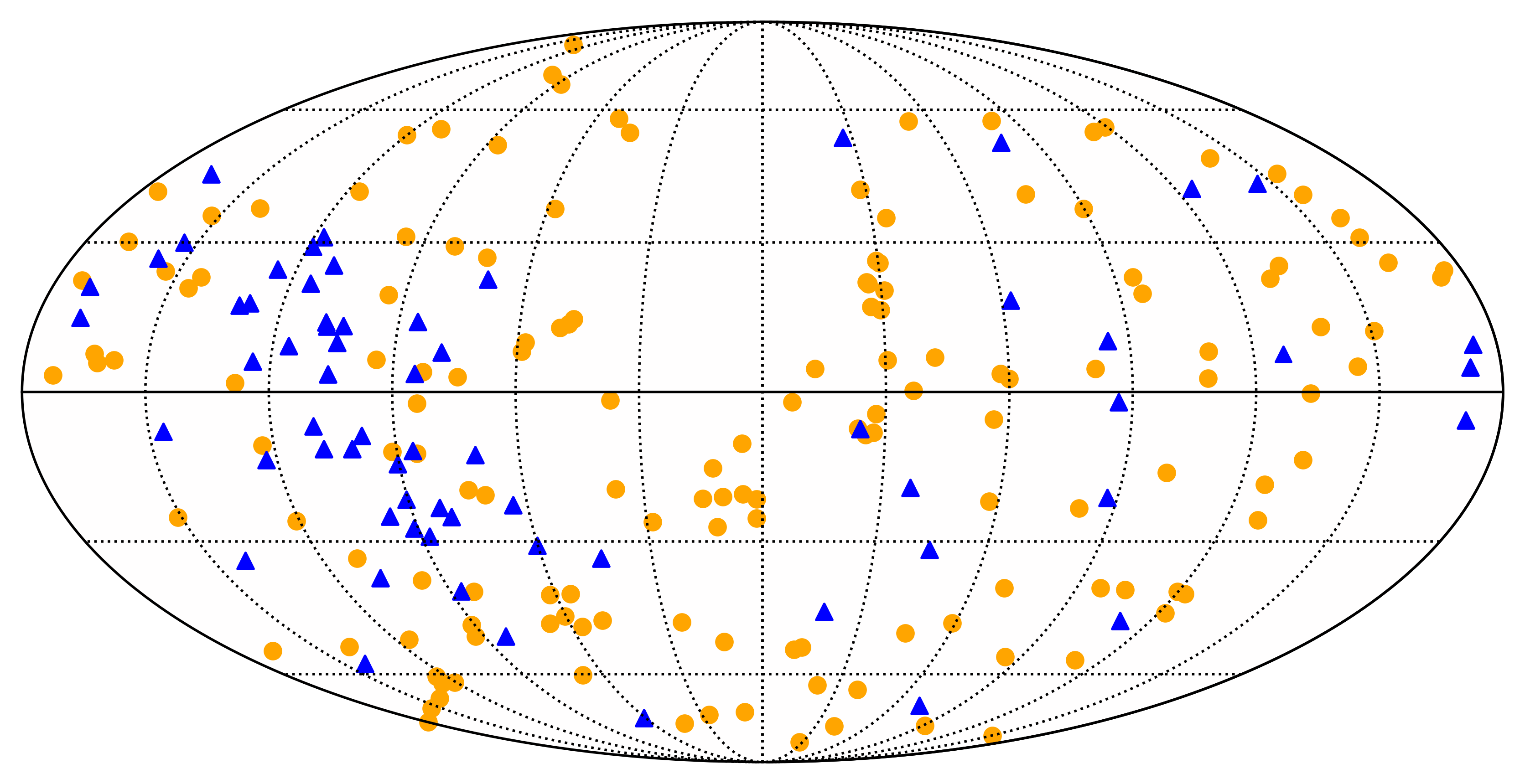}
  \caption{Mollweide projection of our 150 synthetic pulsars, distributed uniformly across the full sky (orange circles).
    For comparison, we also show the observed pulsars from IPTA DR2 (blue triangles)~\cite{Perera:2019sca}.}
  \label{fig:mollview}
\end{figure}

We assume all synthetic pulsars have a common observation start time and are observed at the same cadence.
We randomly populate 150 pulsars isotropically across the full sky, as shown in Fig.~\ref{fig:mollview}.
We also show the pulsar locations from the International PTA (IPTA) data release 2 (DR2)~\cite{Perera:2019sca} for comparison.
For the 100 (50) pulsar analyses, we select a random subset of the 150 (100) pulsars.
The locations of these pulsars are fixed for all analyses involving the same number of pulsars.

We generate the list of TOAs for each pulsar using \texttt{TEMPO2}~\cite{Hobbs:2006cd} and its Python wrapper \texttt{libstempo}.\footnote{https://github.com/vallis/libstempo}
We assume an average observation cadence of 14 days (with small random variations in time for each individual TOA measurement) to create a set of $T_{{\rm obs}}/(14~\text{days})$ number of TOAs for each pulsar.
The TOAs need to be adjusted to account for measurement and instrument error, pulsar intrinsic red noise, and the presence of an isotropic SGWB.

We randomly sample noise parameters from truncated normal, truncated log-normal, or uniform distributions, as listed in Table~\ref{tab:table2}.
For each pulsar $a$, we generate values for instrument errors ($\efac{a}$ and $\equad{a}$) and intrinsic red noise ($A_{\text{RN},a}$ and $\gamma_{\text{RN},a}$).
Furthermore, the $i^{th}$ TOA receives a measurement error ($\sigma_{\text{TOA}, ai}$).
Note that we set the maximum on the ranges for $A_{\text{RN},a}$ and $\gamma_{\text{RN},a}$ so that there is a high probability the SGWB signal will dominate the pulsar intrinsic red noise at lower frequencies where the SGWB is modeled.
These maximum ranges help ensure each synthetic pulsar contributes to a SGWB signal that is distinguishable from the noise of that pulsar.

\begin{table}[t]
  \centering
  \begin{tabular}{|p{2.4cm}|p{1.75cm}|p{1.75cm}|p{1.8cm}|}
    \hline
    Noise input & Mean & Std. dev. & Range \\
    \hline
    $\sigma_{\text{TOA}, ai}$ [ns] & 100 & 30 & $[9, 601]$ \\
    $\efac{a}$ & 1.0 & 0.05 & $[0.5,5.0]$  \\
    $\log_{10}(\equad{a}~[\text{s}])$ & $-8.5$ & 0.1 & $[-11, -6]$ \\
    \hline
    $\log_{10} A_{\text{RN},a}$ & $-16$ & 1 & $[-18,-14.7]$ \\
    \cline{2-3}
    $\gamma_{\text{RN},a}$ & \multicolumn{2}{|c|}{Uniform Distribution} & $[1,5]$ \\
    \hline
  \end{tabular}
  \caption{Pulsar white- and red-noise parameters for generating mock PTA data.
    The first column denotes the noise parameter of interest, with units indicated where relevant.
    In order to randomly generate values of these parameters for each pulsar $a$ and/or TOA $i$, we sample most of the parameters using a truncated normal distribution with a mean, standard deviation, and range given in the remaining columns.
    We sample $\gamma_{\text{RN}}$ using a uniform distribution.
    The values for $\sigma_{\text{TOA}, ai}$ correspond to optimistic-quality pulsars in Ref.~\cite{Xin:2020owo}.}
  \label{tab:table2}
\end{table}

We inject a SGWB signal into the mock TOAs, as described in Ref.~\cite{Chamberlin:2014ria} and we briefly summarize in Appendix~\ref{app:seeds}.
We use the frequency power spectrum in Eq.~\eqref{eq:GWpowerlaw} with spectral index $\gamma_{\text{gw}} = 13/3$ and amplitude $A_{\text{gw}}=2 \times 10^{-15}$, which corresponds to the lower end of the reported common red-noise process~\cite{Antoniadis:2022pcn, NANOGrav:2020bcs, Goncharov:2021oub, Chalumeau:2021fpz}.

There are additional pulsar properties we must specify in order to use the full suite of PTA analysis software (i.e., pulsar spin frequency, spin-down rate, parallax, and dispersion measure).
We populate the values of these properties by randomly sampling from empirical distributions that we create using IPTA DR2 pulsar attributes from Ref.~\cite{Perera:2019sca}.
These additional properties do not otherwise play a role in or impact our study, so we do not discuss them further.

Each of our nine mock PTA datasets are generated with the same set of pseudorandom number generator seeds used to inject the white-noise, red-noise, and SGWB signals.
We use the same set of seeds in order to focus on differences arising from varying $T_{{\rm obs}}$ and $N_p$, without the confounding effects of different realizations of pseudorandom number generation.
To ensure the results of our mock datasets are not driven by outlier seeds, we perform a realization study, described in Appendix~\ref{app:seeds}.
Each realization analysis involves creating mock data with 100 different sets of seeds to produce 100 different realizations, and all other aspects of the analysis remain the same.
We find that the main qualitative results of this paper are unchanged with different mock PTA dataset realizations.

%%%%%%%%%%%%%%%%%%%%%%%%%%%%%%%%%%%%%%%%
\subsection{Analysis methods}

In order to implement the harmonic analysis presented in Sec.~\ref{sec:analysis-harmonic}, we use \texttt{ENTERPRISE}~\cite{enterprise} and \texttt{enterprise-extensions}~\cite{enterprise-extensions} to calculate the likelihood in Eq.~\eqref{eq:Likelihood}.
We modify \texttt{enterprise-extensions} to include the angular correlation from Eq.~\eqref{eq:LegendreCoefficient} with the Legendre coefficients as model parameters.
We use \texttt{PTMCMCSampler}~\cite{justin_ellis_2017_1037579} to perform Markov chain Monte Carlo (MCMC) sampling to determine parameter posterior distributions from each of our nine mock datasets.
The parameters and their prior ranges are listed in Table~\ref{tab:table3}.
The lower limit of the Legendre coefficient prior range comes from the requirement that the angular power spectrum given by Eq.~\eqref{eq:SGWBAngularPS} is a strictly positive quantity.
The upper limit of the Legendre coefficient prior range comes from the requirement that the SGWB two-point correlation function given by Eq.~\eqref{eq:GWCovariance} must be positive definite.

\begin{table}[t]
  \centering
  \begin{tabular}{|p{2.1cm}|p{3.1cm}|p{2.8cm}|}
    \hline
    Parameter & \multicolumn{2}{|c|}{MCMC prior range}\\
    \cline{2-3}
    & Single-pulsar analysis & Harmonic analysis \\
    \hline
    $\log_{10} A_{\text{gw}}$ & --- & [-18,-14]\\
    $\gamma_{\text{gw}}$ & --- & Fixed at 13/3\\
    $c_{2}$ through $c_{8}$ & --- & [0,1]\\
    \hline
    $\efac{a}$ & $[0.01,10]$ & Fixed to best fit \\
    $\log_{10}\equad{a}$ & $[-12,-5]$ & Fixed to best fit\\
    \hline
    $\log_{10} A_{\text{RN},a}$ & [-20,-11] & [-20,-11] \\
    $\gamma_{\text{RN},a}$ & [0,7] & [0,7]\\
    \hline
  \end{tabular}
  \caption{Uniform prior ranges for MCMC single-pulsar and harmonic analyses.
    The single-pulsar analysis does not include angular correlations or the frequency power spectrum for a SGWB.
    The white-noise parameters $\efac{a}$ and $\log_{10}\equad{a}$ in the harmonic analysis are fixed to their maximum-likelihood values obtained from the single-pulsar analysis.}
  \label{tab:table3}
\end{table}

Following standard PTA methods~\cite{Antoniadis:2022pcn, NANOGrav:2020bcs, Goncharov:2021oub, Chalumeau:2021fpz}, we perform a single-pulsar noise analysis for a given mock PTA dataset before running the main analysis to extract SGWB properties.
The single-pulsar analysis involves four parameters for each pulsar: the two white-noise instrument error parameters ($\efac{a}$ and $\equad{a}$) and the two pulsar intrinsic red-noise parameters ($A_{\text{RN},a}$ and $\gamma_{\text{RN},a}$).
SGWB parameters are not included: there are no angular correlations for a single pulsar, and the common red-spectrum process of the SGWB cannot be distinguished from the pulsar intrinsic red noise.
Thus, the covariance matrix $\mathcal{C}$ in Eq.~\eqref{eq:Likelihood} does not include any cross-correlations between pulsars for the single-pulsar analysis.
Note that we use the same number of frequency components to analyze the pulsar intrinsic red noise as those used in the main harmonic analysis.

\begin{figure*}[t]
  \centering
  \includegraphics[width=0.98\textwidth]{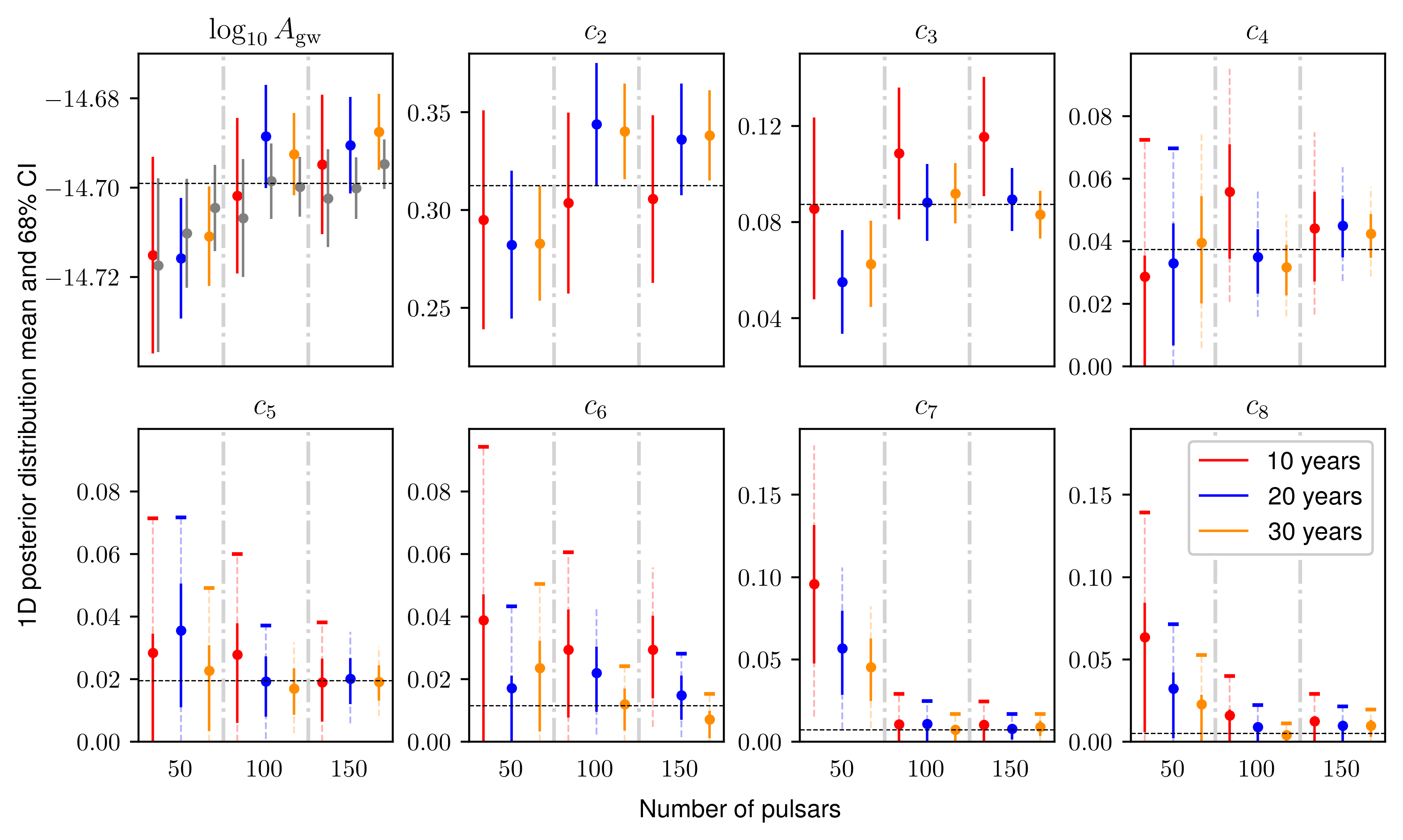}
  \caption{Mean and 68\% CI of the marginalized 1D posterior distributions for $\log_{10} A_{\text{gw}}$ and Legendre multipole coefficients for $\ell=2$ through $\ell=8$ for all nine harmonic analyses.
    Each panel separates the analyses for different numbers of pulsars with a vertical dot-dashed line, and analyses for different observation times are shown in different colors, as indicated in the legend.
    For multipoles $\ell=4$ through $\ell=8$, we also show the 95\% CI as dashed lines.
    When the parameter is consistent with zero in the 95\% CI, we show a line cap at the upper 95\% limit.
    For each parameter, the injected value is shown as a horizontal dashed, black line.
    For $\log_{10} A_{\text{gw}}$, the gray points next to each colored point represent the corresponding HD analyses, in which the Legendre coefficients are fixed to their injected values.
  }
  \label{fig:CIplot}
\end{figure*}

From the single-pulsar analysis, we obtain maximum-likelihood values for $\efac{a}$ and $\equad{a}$.
These values for $\efac{a}$ are generally consistent with their injected values (i.e., the values we use to generate the signals for our mock data), while the values for $\equad{a}$ are poorly recovered, because the injected values of $\equad{a}$ are subdominant to $\sigma_{\text{TOA},ai}$ [cf.\ Eq.~\eqref{eq:WNvariance}].
However, in turn, the poorness of the recovery has little impact on our main analysis.

For the full PTA analyses, we fix the values of $\efac{a}$ and $\equad{a}$ to their maximum-likelihood values obtained from the single-pulsar analysis.
The pulsar intrinsic red-noise parameters and the SGWB parameters, including the Legendre coefficients, are varied according to Table~\ref{tab:table3} in the MCMC analysis.

We run multiple MCMC chains in parallel to reduce processing time when analyzing a given mock dataset; we do not employ parallel tempering.
We combine sampling chains after removing a 25\% burn-in to create a single final chain.
We use the Gelman-Rubin $R$-statistic~\cite{Gelman:1992zz} as a measure of chain convergence and require $R-1<0.1$ for all SGWB parameters.
We also require a minimum combined total of $3.5 \times 10^6$ samples prior to removing the burn-in.
The sample chains are thinned by a factor of 10, which is the default for \texttt{PTMCMCSampler}.
The initial values of the Legendre coefficients, randomly drawn from their prior ranges, are scaled by the total number of Legendre coefficients to prevent MCMC from getting stuck at the initial sample point.

As a point of comparison, we also perform an HD analysis by fixing the values of the Legendre coefficients to their theoretical values in Eq.~\eqref{eq:LegendreCoefficientscale}.
Fixing the coefficients is equivalent to using the HD correlation function given by Eq.~\eqref{eq:HDcurve}, modulo any effects due to truncating the angular power spectrum at $\ell=8$.
The only SGWB parameter in these HD analyses is $\log_{10} A_{\text{gw}}$, and we use the same prior range given in Table~\ref{tab:table3} for the harmonic analysis.

%%%%%%%%%%%%%%%%%%%%%%%%%%%%%%%%%%%%%%%%
\subsection{Results}
\label{sec:results}

\begin{figure}[t]
  \centering
  \includegraphics[width=0.48\textwidth]{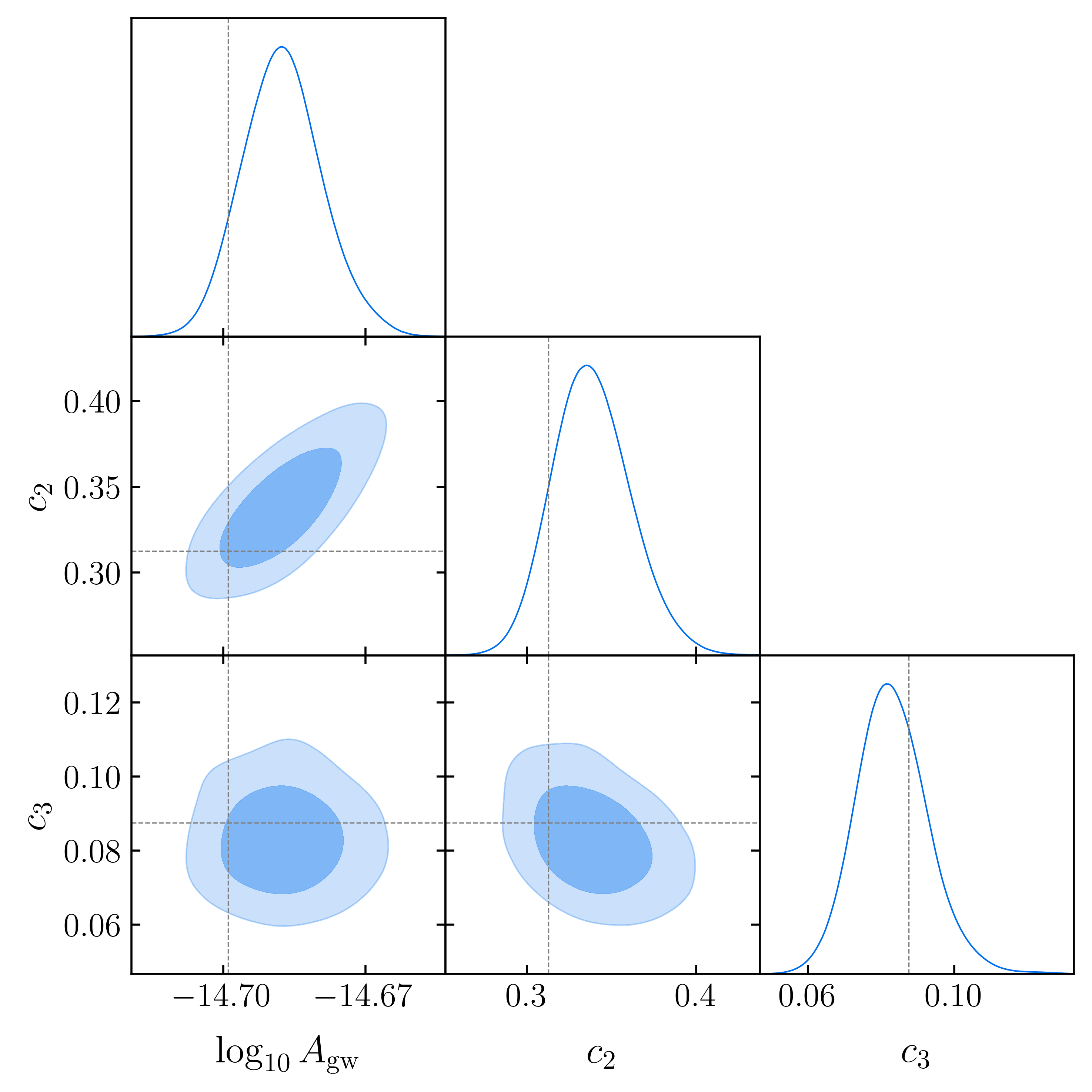}
  \caption{Corner plot of the marginalized 1D and 2D posterior distributions for $\log_{10} A_{\text{gw}}$, $c_2$, and $c_3$ from the harmonic analysis with 150 pulsars observed for 30 years.
    The vertical and horizontal dashed lines indicate the injected values.
    There is a moderately strong positive correlation between $\log_{10} A_{\text{gw}}$ and $c_2$ and a weak negative correlation between $c_2$ and $c_3$.
    The corner plots for all SGWB parameters in all nine harmonic analyses are provided in Appendix~\ref{app:corner}.
  }
  \label{fig:Corner}
\end{figure}

\begin{figure}[t]
  \centering
  \includegraphics[width=0.48\textwidth]{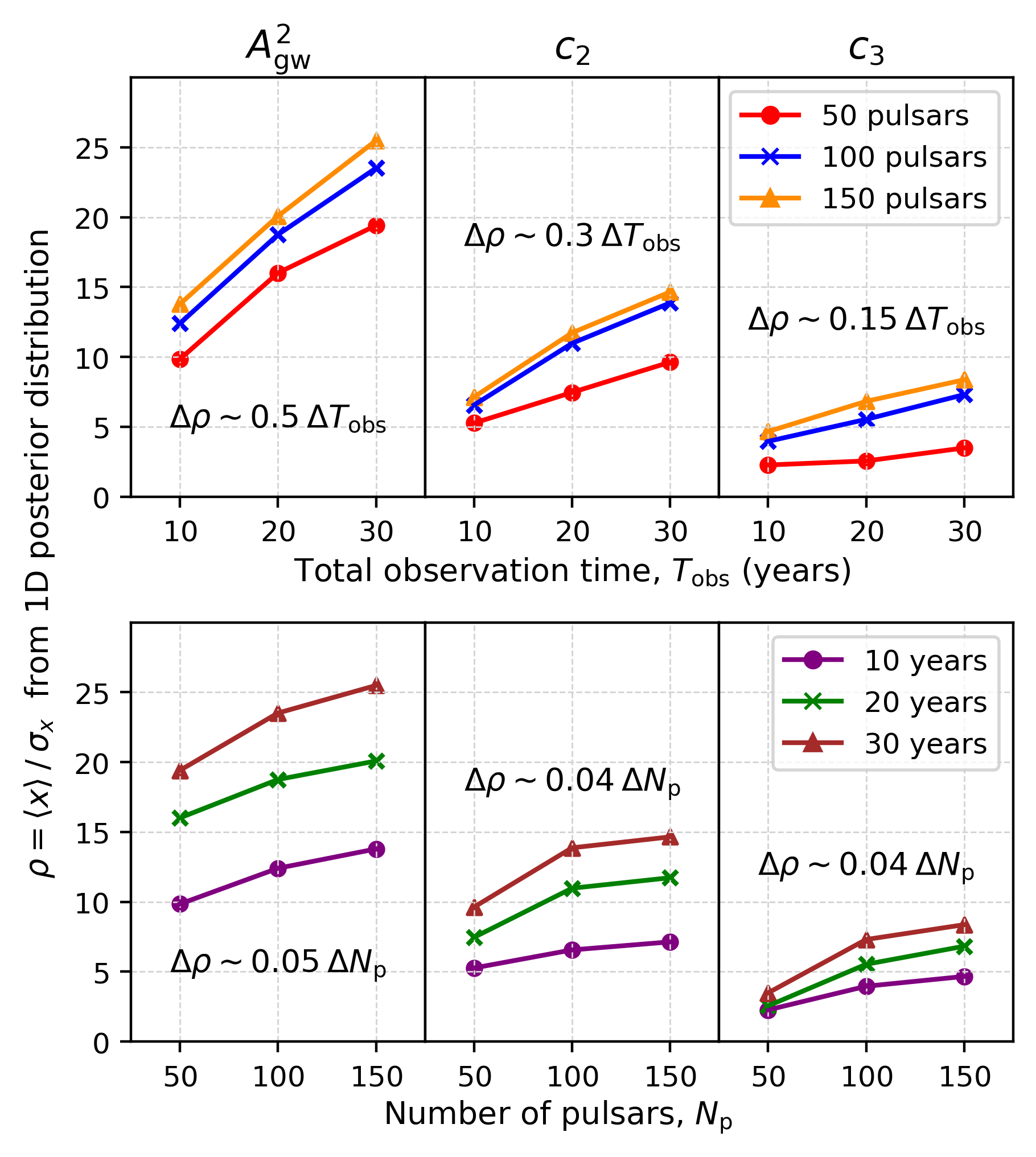}
  \caption{Ratio of the mean $\langle x \rangle$ to one-half the width of the 68\% credible interval $\sigma_x$ for the marginalized 1D posterior distribution of the parameter $x \in \{ A_{\text{gw}}^2, c_2, c_3 \}$, indicated at the top of each column of panels, for all nine harmonic analyses.
    We plot the ratios in the top and bottom rows as a function of observation time and number of pulsars, respectively.
  }
  \label{fig:SNRplot}
\end{figure}

\begin{figure}[t]
  \centering
  \includegraphics[width=0.48\textwidth]{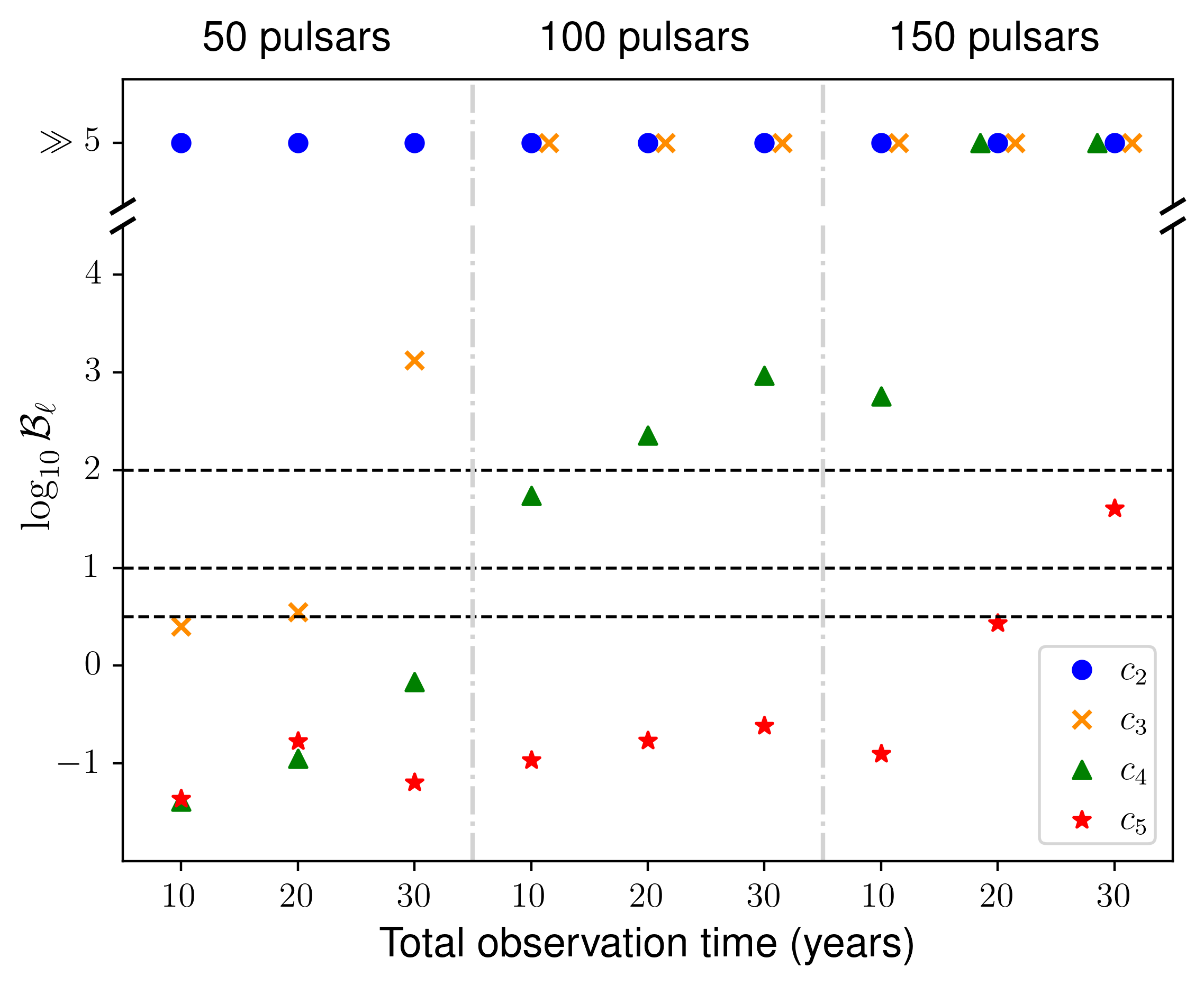}
  \caption{Savage-Dickey Bayes factors $\mathcal{B}_{\ell}$ for $\ell=2$ to $\ell=5$ for all nine harmonic analyses.
    The Bayes factor provides a measure of the evidence for including a Legendre coefficient $c_{\ell}$ in the harmonic analysis.
    The horizontal dashed lines show the commonly used separation of the Bayes factor strength of evidence: $\log_{10}\mathcal{B}_{\ell}>2$ is decisive, $1<\log_{10}\mathcal{B}_{\ell}<2$ is strong, $0.5<\log_{10}\mathcal{B}_{\ell}<1$ is substantial, and $\log_{10}\mathcal{B}_{\ell}<0.5$ is no evidence.
    Coefficients for $\ell>5$ are not included, since we generally observe $\log_{10}\mathcal{B}_{\ell} \ll 0.5$ for these multipoles.
    For well-measured coefficients, the Bayes factor is very large, which we denote as $\log_{10} \mathcal{B}_\ell \gg 5$.
    We include a small horizontal offset for a few points to prevent overlap.
  }
  \label{fig:SDplot}
\end{figure}

\begin{figure*}[t]
  \centering
  \includegraphics[width=0.48\textwidth]{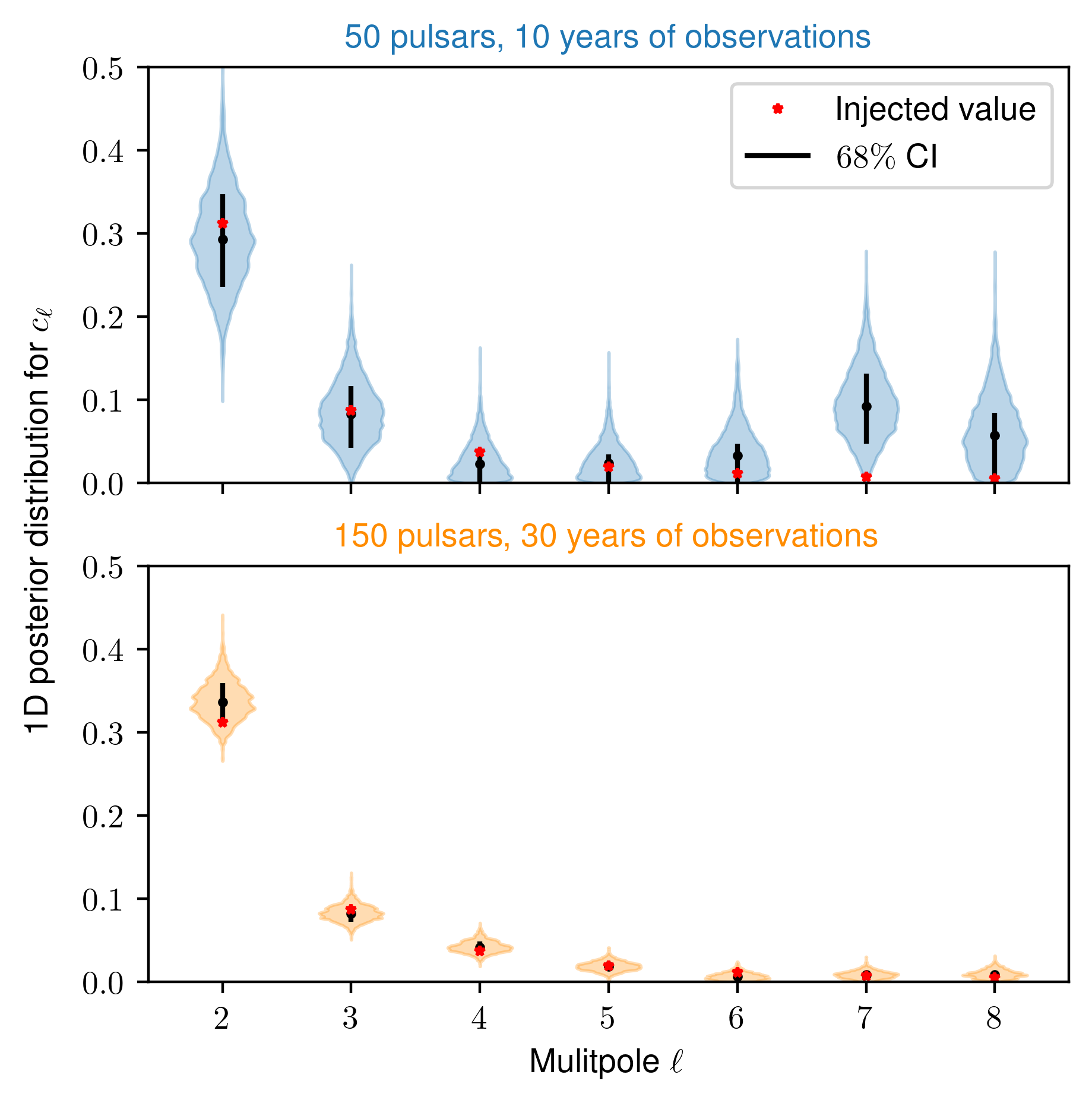}
  \includegraphics[width=0.48\textwidth]{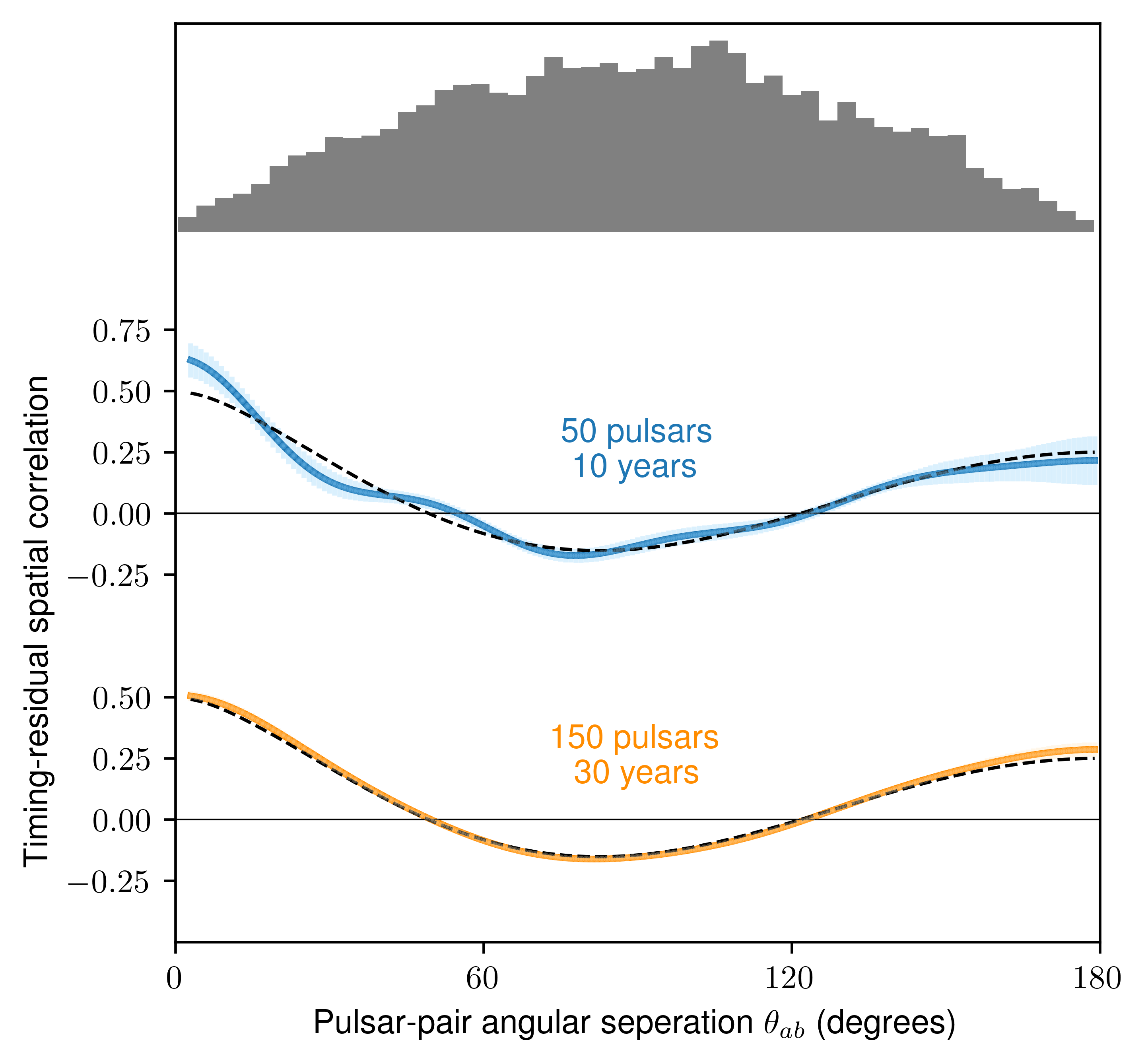}
  \caption{Left: violin plots of the marginalized 1D posterior distributions for the Legendre coefficients from the two harmonic analyses with the lowest (top panel) and highest (bottom panel) overall multipole evidence.
    We show the mean (black dot) and 68\% CI (black line) for each posterior distribution, along with the injected values of the Legendre coefficients (red stars) from Eq.~\eqref{eq:LegendreCoefficientscale}, corresponding the HD angular correlation function.
    Right: reconstructed angular correlation function using the multipole coefficients shown in the left panels, using 100 angular-separation bins.
    The solid blue and orange curves represent the mean value within each angular-separation bin, and the error bars show the $1\sigma$ deviation of each bin.
    The dashed black line is the HD curve.
    At the top of this panel, we show the histogram of angular separations for the 150 synthetic pulsars we use for our harmonic analyses.
  }
  \label{fig:BCWCplot}
\end{figure*}

For our nine harmonic analyses, we show the mean and 68\% credible interval (CI) of the marginalized 1D posterior distributions for the SGWB parameters in Fig.~\ref{fig:CIplot}.
We provide the corner plots for all SGWB parameters of our nine harmonic analyses in Appendix~\ref{app:corner}.
For the SGWB amplitude, the quadrupole, and multipole $\ell=3$, the posterior distributions can be accurately characterized by the mean and 68\% CI (as shown in Appendix~\ref{app:corner}).
For multipoles $\ell=4$ through $\ell=8$, we also show in Fig.~\ref{fig:CIplot} the 95\% CI with dashed error bars.
When the multipole's posterior distribution cannot be distinguished from zero with a 95\% CI, we place a line cap at the 95\% CI upper limit.
The injected value is plotted for each parameter as a dashed black horizontal line.
We also show the results of our HD analyses, in which the Legendre coefficients are fixed to their injected values from Eq.~\eqref{eq:LegendreCoefficientscale}, for $\log_{10} A_{\text{gw}}$ in Fig.~\ref{fig:CIplot}.
The HD analysis results are plotted in gray, immediately to the right of the analogous harmonic analysis.

For detected parameters (i.e., parameters whose 95\% CI does not include zero), Fig.~\ref{fig:CIplot} shows the spread of the posterior distribution decreases and the mean tends toward the injected value of the parameter as we increase $T_{\rm obs}$ and $N_p$, which is intuitively reasonable.
For parameters that are not detected, their 95\% CI upper limits decrease.
The relatively large 68\% CI that we see with $\ell=7$ for the 50 pulsar analyses is realization dependent and within the expected range of fluctuations for parameters that are not detected, as we discuss in Appendix~\ref{app:seeds}.

For the 50 pulsar analyses, the posterior of $\log_{10} A_{\text{gw}}$ is biased low relative to the injected value of $\log_{10} A_{\text{gw}}$.
The bias comes from modeling pulsar intrinsic red noise, which absorbs some of the power from the injected SGWB signal~\cite{Hazboun:2020kzd}.
The magnitude of this bias decreases as we increase $T_{\rm obs}$ and $N_p$, because the SGWB amplitude in the cross-correlations of the timing data helps to distinguish the SGWB signal from the pulsar intrinsic red noise.

Fig.~\ref{fig:CIplot} shows the 68\% credible interval of the posterior distribution for $\log_{10}A_{\text{gw}}$ has approximately the same scale between the harmonic analysis and the corresponding HD analysis, which means adding multipoles in our harmonic analysis has a minimal affect on our ability to recover the SGWB amplitude.
However, we observe in Fig.~\ref{fig:CIplot} that the mean of the posterior distribution for $\log_{10}A_{\text{gw}}$ in the harmonic analysis trends toward larger values than the corresponding HD analysis as we increase $T_{\rm obs}$ and $N_p$.
This effect is caused by a moderately strong positive correlation between $\log_{10} A_{\text{gw}}$ and $c_2$, which we show in the corner plot of Fig.~\ref{fig:Corner}.
We can see in Fig.~\ref{fig:CIplot} that $c_2$ is trending high relative to its injected value as we increase $T_{\rm obs}$ and $N_p$, so the positive correlation is causing a similar trend in $\log_{10} A_{\text{gw}}$ for the harmonic analyses.
We show in Appendix~\ref{app:seeds} that detected parameters generally recover the injected value within $1\sigma$, with fluctuations that are realization dependent and consistent with an ergodic process.
Therefore, the trend we observe for $\log_{10} A_{\text{gw}}$ and $c_2$ is realization dependent.

In Fig.~\ref{fig:Corner} we also observe a weak negative correlation between $c_2$ and $c_3$.
The weak negative correlation between multipoles is not due to the harmonic mode coupling from a finite number of pulsars~\cite{Roebber:2019gha}, because the correlations get stronger as the number of pulsars increases, as can be seen in the corner plots of Appendix~\ref{app:corner}.
We expect to observe some correlations between SGWB parameters because of their functional relationship in the angular power spectrum.

In Fig.~\ref{fig:SNRplot} we plot $\rho_x \equiv \left<x \right> / \sigma_x$ where $\left<x \right>$ is the mean of the marginalized 1D posterior distribution and $\sigma_x$ is one-half the width of the 68\% credible interval for $x \in \{ A_{\text{gw}}^2, c_2, c_3 \}$.
Our calculated parameter $\rho_x$ is a measure of the significance of parameter $x$ relative to its uncertainty from its posterior distribution.
We obtain the distribution for $A_{\text{gw}}^2$ by transforming the posterior distribution of $\log_{10} A_{\text{gw}}$.
We choose these three parameters to calculate $\rho_x$ because they are detected in all analyses, as shown in Fig.~\ref{fig:CIplot}.

The top row of Fig.~\ref{fig:SNRplot} shows the nine harmonic analyses plotted as a function of $T_{\rm obs}$, while the bottom row shows these same nine analyses plotted as a function of $N_p$.
We include in Fig.~\ref{fig:SNRplot} the linear scaling relationship between a change in $\rho_x$ versus a change in $T_{\rm obs}$ (top row) or a change in $N_p$ (bottom row), which we obtain by linear regression.
The scaling of $A_{\text{gw}}$ with both $T_{\rm obs}$ and $N_p$ roughly matches the scaling shown in Figs.~9 and~11, respectively, of Ref.~\cite{vanHaasteren:2008yh} once we rescale $\rho_{A_{\text{gw}}^2}$ by a factor of $1/2$ to account for the fact that we are considering $A_{\rm gw}^2$, whereas Ref.~\cite{vanHaasteren:2008yh} considers $A_{\rm gw}$.
The difference is largest for the scaling with $T_{\rm obs}$.
This difference may be due to the fact our PTA properties differ and that we perform a full MCMC, whereas Ref.~\cite{vanHaasteren:2008yh} fixes all other parameters to their maximum-likelihood value.

Increasing $T_{{\rm obs}}$ has a larger affect on $\rho_x$ than increasing $N_{p}$.
This effect occurs, because increasing $T_{\rm obs}$ increases the number of low-frequency harmonics where the strength of the SGWB signal in the autocorrelations dominates the white noise~\cite{Siemens:2013zla, Romano:2020sxq}.
For our Bayesian analysis approach, we model a finite number of harmonics of $f_{L}=1/T_{{\rm obs}}$, as discussed in Sec.~\ref{sec:analysis-harmonic}.
By using the average value of our injected white noise, we find for $T_{{\rm obs}}=10$~years only the first few harmonics of $f_L$ are in the regime where the SGWB signal is dominant, while for $T_{{\rm obs}}=30$~years the first ten frequency harmonics are in this regime.
So increasing $T_{{\rm obs}}$ provides more frequency bins where the strength of the SGWB signal dominates the white noise in the autocorrelations, thereby reducing the spread of the SGWB amplitude distribution and increasing $\rho_{A_{\text{gw}}^2}$.

We use the Savage-Dickey approach~\cite{10.2307/2958475} to calculate a Bayes factor for each multipole in our harmonic analyses, which is a measure of the evidence for the multipole in our model.
The calculation methodology is provided in Appendix~\ref{app:savage-dickey}.
Fig.~\ref{fig:SDplot} shows the Savage-Dickey Bayes factors for multipoles up to $\ell=5$ in the nine harmonic analyses.
General trends in Fig.~\ref{fig:SDplot} are consistent with basic expectations; i.e., evidence for including higher multipoles increases as $T_{\rm obs}$ and $N_p$ increase due to increasing $\rho_{A_{\text{gw}}^2}$.
We see there is decisive evidence for the quadrupole in all harmonic analyses.
As $\rho_{A_{\text{gw}}^2}$ increases, evidence for $\ell=3$ and $\ell=4$ becomes decisive; for 150 pulsars observed for 30 years, we see strong evidence for $\ell=5$.
Multipoles $\ell>5$ are not shown in Fig.~\ref{fig:SDplot}, because they do not show trending evidence (relative to increasing $T_{\rm obs}$ and $N_p$) for being included in the model.

The left plot in Fig.~\ref{fig:BCWCplot} shows the multipole marginalized 1D posterior distributions from the harmonic analyses that have the lowest (top) and highest (bottom) overall multipole evidence.
These violin plots show the improvement in both the mean (relative to the injected value) and the spread of the posterior distributions as we increase $T_{\rm obs}$ and $N_p$.
The right plot in Fig.~\ref{fig:BCWCplot} shows the reconstructed angular correlation function from the multipoles shown in the left plot.
The histogram at the top shows the number of pulsar pairs binned by angular separation.
To reconstruct the angular correlation function, the values of the Legendre coefficients at each MCMC chain step are inserted into Eq.~\eqref{eq:LegendreCoefficient} at 100 different angular separation bins.
The solid blue and orange curves represent the mean value within each angular separation bin, and error bars are the $1\sigma$ deviations.
The top blue plot shows that even when multipoles $\ell>3$ have little-to-no evidence for being included in the model, the HD angular correlation function can be fairly accurately reconstructed because of the strong quadrupolar dependency and the sharp dropoff in the angular power spectrum as $\ell$ increases.
The bottom orange plot show how accurately the HD curve can be reconstructed when we have evidence for multipoles up to $\ell=5$.

Lastly, we provide an example of how our harmonic analysis formalism can be used to constrain scenarios that modify GR.
We use the harmonic analysis with 150 pulsars and 30 years of observations to determine the sensitivity to subluminal GW propagation.
The tensor-mode detector response function $F_\ell$ from Table I of Ref.~\cite{Qin:2020hfy} is a function of the GW subluminal group velocity $v_{\rm gw}$, where we assume a single GW phase velocity with no frequency dependence [to ensure the factorization in Eq.~\eqref{eq:GWCovariance} holds].
From Eq.~\eqref{eq:LegendreCoefficientGeneral}, we find $c_{\ell+1} / c_{\ell}$ as a function of $v_{\rm gw}$, where our use of ratios is to ensure scaling of the autocorrelation cancels.
The 95\% lower limit of the 1D posterior distributions of $c_3/c_2$, $c_4/c_3$, and $c_5/c_4$ from our harmonic analysis provides corresponding 95\% lower limits on the GW group velocity of $v_{\rm gw}/c=0.95$, $v_{\rm gw}/c=0.98$, and $v_{\rm gw}/c=0.96$, respectively.
It is interesting to note that, as the group velocity decreases the power spectrum at $\ell>2$ decreases so that even though the best-measured multipoles are the quadrupole and octupole, our strongest constraint on $v_{\rm gw}$ comes from the ratio of the hexadecapole to the octupole.
Note that even though ground-based measurements of GW place tight constraints on the speed of propagation of GWs~\cite{LIGOScientific:2016aoc,LIGOScientific:2017zic}, this speed may be different in the frequency range accessible to PTAs~\cite{deRham:2018red}.

%%%%%%%%%%%%%%%%%%%%%%%%%%%%%%%%%%%%%%%%%%%%%%%%%%%%%%%%%%%%%%%%%%%%%%%%%%%%%%%
\section{Conclusions}
\label{sec:conclusions}

In this paper we demonstrate the capabilities and limitations of a harmonic analysis for mock PTA timing data that includes an isotropic SGWB.
We model the first seven nonzero multipoles (i.e., $\ell=2$ through $\ell=8$) of the SGWB angular power spectrum using a Legendre power series representation of the angular correlation function.
We perform our harmonic analyses on mock PTA datasets with different numbers of pulsars (50, 100, and 150) and pulsar observation times (10, 20, and 30 years) for pulsars uniformly distributed across the sky.

We find decisive evidence of the quadrupole contribution in the angular power spectrum for all of our harmonic analyses.
For higher multipoles, the general trend we observe is that for 50 pulsars evidence for $\ell = 3$ is strong to decisive, for 100 pulsars evidence for $\ell = 3$ is decisive and $\ell = 4$ is strong to decisive, and for 150 pulsars evidence for multipoles up to $\ell = 4$ is decisive and the evidence for $\ell = 5$ becomes strong when evidence of the SGWB amplitude is sufficiently high.
The first four nonzero multipoles in the angular power spectrum can accurately reconstruct the HD curve (to within 2\% of the spatial correlations integrated over angular separations) due to the sharp dropoff in multipole values as $\ell$ increases.
Therefore, our harmonic analysis approach is a promising tool to help determine angular correlations present in pulsar timing data which includes searching for anisotropies in the SGWB as well as searching for other cosmological and astrophysical sources of GWs.

For $A_{\text{gw}}^2$, $c_2$, and $c_3$ we provide linear scaling relationships as a function of $T_{\rm obs}$ and as a function of $N_{p}$.
We find that increasing $T_{\rm obs}$ has a larger affect than increasing $N_{p}$.
We also compare the scaling of $A_{\text{gw}}^2$ to previous work and find that our scaling with $N_{p}$ is consistent with previous work, but our scaling with $T_{\rm obs}$ is higher by approximately a factor of 2, which we suspect is due to different modeling techniques.

For higher multipoles that are not detected in our model, we can place upper limits that decrease sharply with increasing $T_{\rm obs}$ and $N_p$.
We observe that multipoles are not detected until the theoretical value of the Legendre coefficient for that multipole is less than the posterior standard deviation of $\log_{10}A_{\text{gw}}$.

Most analyses of PTA data assume that the angular correlations between pulsars agree with the standard expectations of an isotropic SGWB and GR.
As the data improve, it will be essential to allow for us to test these assumptions and allow deviations from these standard expectations.
The method we present here provides a flexible parameterization that enables us to more fully explore the implications of any future detection of angular correlations in a PTA, giving us the ability to explore deviations from GR (e.g.\ Ref.~\cite{NANOGrav:2021ini}), anisotropy (e.g.\ Ref.~\cite{Mingarelli:2013dsa}), and systematic errors which cause angular correlations, such as clock errors and shifts in the Solar System barycenter~\cite{Tiburzi:2015kqa}.

  \emph{Note added.}: Recently, PTA collaborations reported evidence of SGWB spatial correlations consistent with HD correlations at varying levels of significance~\cite{NANOGrav:2023gor, EPTA:2023fyk, Reardon:2023gzh, Xu:2023wog}.

%%%%%%%%%%%%%%%%%%%%%%%%%%%%%%%%%%%%%%%%%%%%%%%%%%%%%%%%%%%%%%%%%%%%%%%%%%%%%%%
\begin{acknowledgments}
  We thank Steve Taylor, Jeff Hazboun, and Nima Laal for helpful discussions regarding the PTA software tools used for these analyses.
  We acknowledge the Texas Advanced Computing Center (TACC) at The University of Texas at Austin for providing high-performance computing resources that have contributed to the research results reported within this paper.
  This work used the Strelka Computing Cluster, which is run by Swarthmore College.
  We used GetDist~\cite{Lewis:2019xzd} to calculate probabilities and credible intervals from parameter posterior distributions and to generate corner plots.

  T.L.S. was supported by National Science Foundation Grant No.~2009377, NASA Grant No.~80NSSC18K0728, and the Research Corporation.
  C.M.F.M. was supported in part by the National Science Foundation under Grants No.~PHY-2020265, and No.~AST-2106552.
  The Flatiron Institute is supported by the Simons Foundation.
\end{acknowledgments}

%%%%%%%%%%%%%%%%%%%%%%%%%%%%%%%%%%%%%%%%%%%%%%%%%%%%%%%%%%%%%%%%%%%%%%%%%%%%%%%
\appendix

\section{Realization Study}
\label{app:seeds}

\begin{figure}[t]
  \centering
  \includegraphics[width=0.48\textwidth]{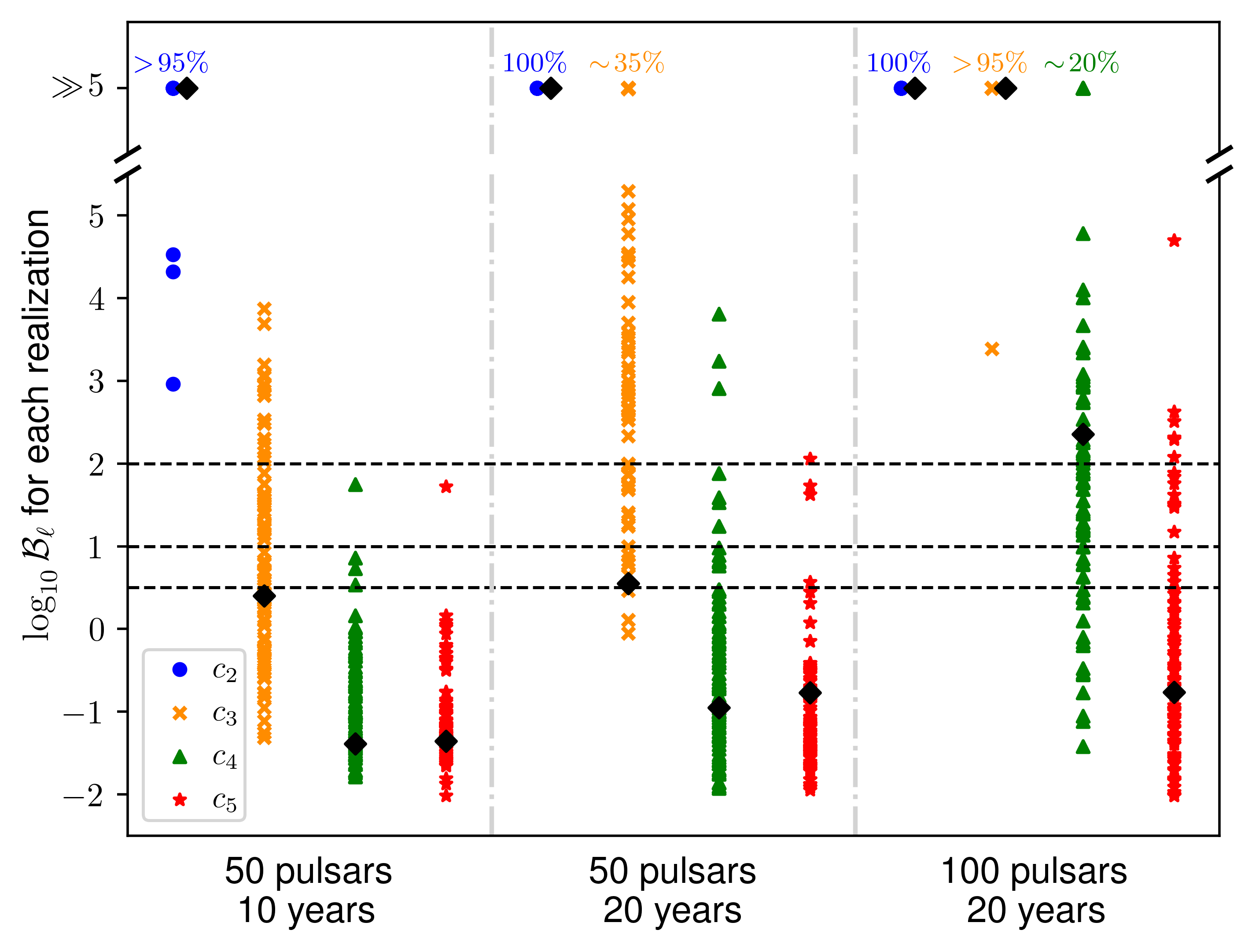}
  \caption{Savage-Dickey Bayes factors for 100 realizations of three mock PTA datasets: 50 pulsars observed over 10 years, 50 pulsars observed over 20 years, and 100 pulsars observed over 20 years.
    The Bayes factors for the realization presented in Sec.~\ref{sec:results} are plotted as black diamonds.
    For cases in which a multipole coefficient has a very large Bayes factor for multiple realizations, we indicate the percentage of realizations with $\log_{10}\mathcal{B} \gg 5$.
    See Fig.~\ref{fig:SDplot} for a further explanation of the plot features.
  }
  \label{fig:SDseeds}
\end{figure}

\begin{figure*}[t]
  \includegraphics[width=0.98\textwidth]{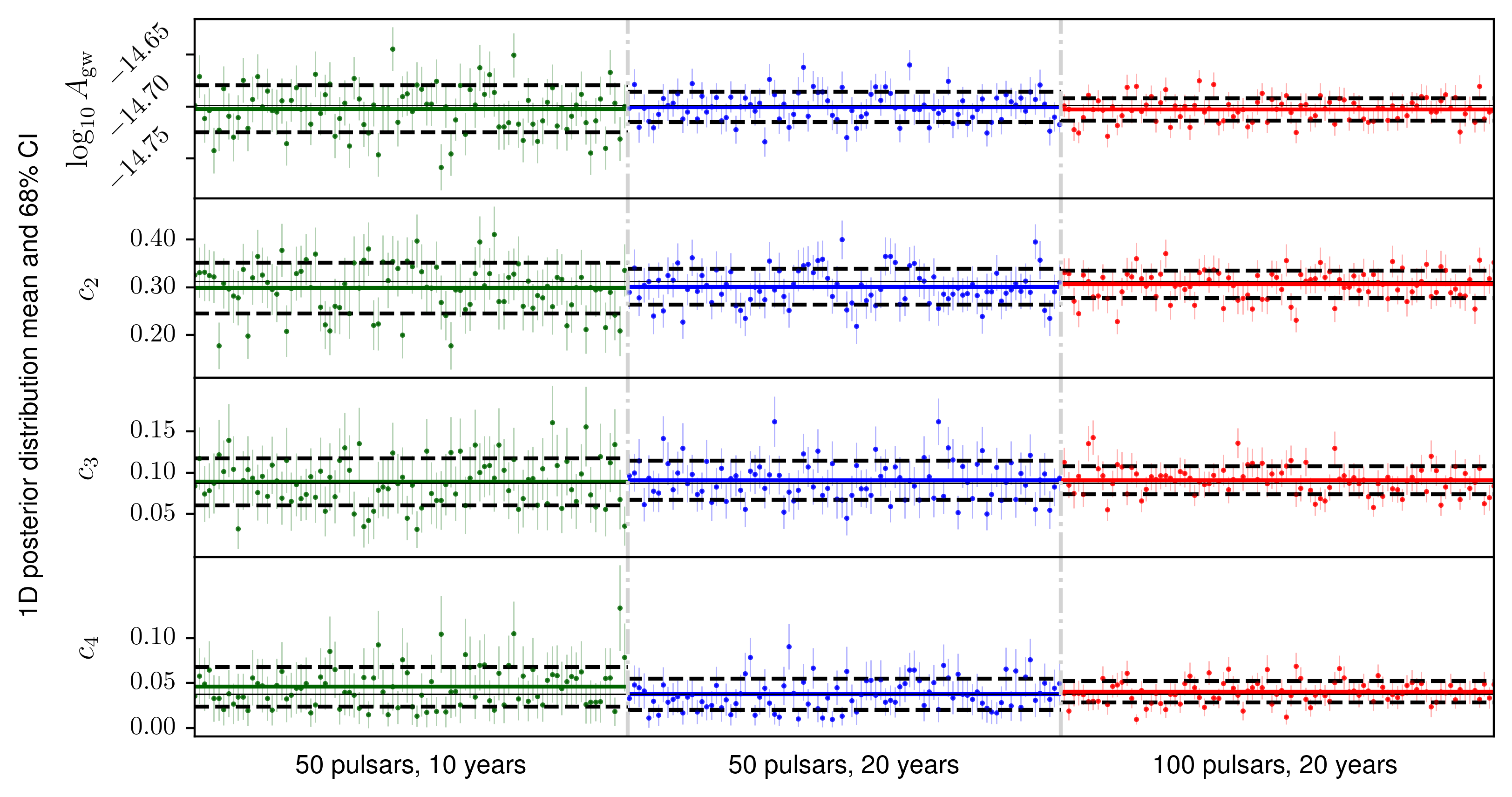}
  \caption{Realization-dependent mean and 68\% CI for parameters $\log_{10} A_{\text{gw}}$, $c_2$, $c_3$, and $c_4$.
    Individual realizations are plotted along the $x$-axis.
    The datasets used for the realization analyses are separated by color and by the gray vertical dot-dashed lines.
    The colored horizontal solid lines represent the averaged mean of the 1D posterior distributions for all realizations of a given dataset, and the black horizontal dashed lines show the standard deviation of the realization means.
    The injected value of each parameter is plotted as a solid horizontal black line.
  }
  \label{fig:CIseeds}
\end{figure*}

In this appendix we present the methodology and results of our realization study in which we vary the pseudorandom number generator seeds.
We choose three different mock PTA datasets to perform these analyses: 50 pulsars observed for 10 years, 50 pulsars observed for 20 years, and 100 pulsars observed for 20 years.
We choose these datasets so that we can observe the differences in realizations on datasets that have different values of $T_{\rm obs}$ and $N_p$.
In the following subsections, we discuss the methodology used to perform the realization study, provide the realization study results, and compare the realization variance of the quadrupole to predicted variance from Refs.~\cite{Roebber:2016jzl, Allen:2022dzg, Allen:2022ksj}.

%%%%%%%%%%%%%%%%%%%%%%%%%%%%%%%%%%%%%%%%
\subsection{Realization study methodology}
\label{sec:seed-method}

We generate 100 realizations for each of our three selected mock PTA dataset by varying the pseudorandom number generator seeds used to inject white noise, pulsar intrinsic red noise, and the SGWB signal into the timing data.
All other analysis methods are the same as those discussed in Secs.~\ref{sec:analysis-harmonic} and~\ref{sec:analysis}.
We provide a brief summary of how pseudorandom number generator seeds inject signals into the timing data using \texttt{TEMPO2} and its Python wrapper \texttt{libstempo}.

For white noise, we first apply TOA measurement error $\sigma_{\text{TOA}, ai}$ to the $i^{th}$ TOA for pulsar $a$, as we discuss in Sec.~\ref{sec:mock}.
Instrument errors $\efac{a}$ and $\equad{a}$, which are the same for all realization analyses of a given mock PTA dataset,  adjust the total white noise using Eq.~\eqref{eq:WNvariance}.
To make these injected white-noise signals Gaussian, $\efac{a}$ and $\equad{a}$ are scaled by random numbers generated from a standard normal distribution based on specified pseudorandom number generator seeds.
We use two different pseudorandom number generator seeds to inject the white noise: one generator seed for the scaling of $\efac{a}$ and one generator seed for the scaling of $\equad{a}$.

We inject intrinsic red noise for pulsar $a$ using the power-law model given by Eq.~\eqref{eq:RNpowerlaw} with values of $A_{\text{RN},a}$ and $\gamma_{\text{RN},a}$ which are the same for all realization analyses of a given mock PTA dataset.
The intrinsic red-noise signal is created in the time domain and added directly to the TOAs.
The first ten harmonics of $1/T_{{\rm obs}}$ are used in a sine-cosine basis representation of the intrinsic red-noise timing signal, as described in Sec.~\ref{sec:analysis-harmonic}.
The amplitudes of the 20 basis functions (ten sine functions and ten cosine functions) are independently scaled by random numbers generated from a standard normal to make the signal Gaussian and then added to the pulsar TOA.
We use a single pseudorandom number generator seed for the scaling applied to the time-domain sine-cosine basis function amplitudes of the intrinsic red noise.

The SGWB signal is injected in the frequency domain using the Fourier transform of Eq.~\eqref{eq:GWCovariance} and performing a Cholesky decomposition of the angular correlation function, as described in Ref.~\cite{Chamberlin:2014ria}.
The Cholesky decomposition provides a matrix $M$ with elements satisfying $\left(M\:M^\dag \right)_{ab} = \zeta(\cos{\theta_{ab}})$, where $\zeta(\cos{\theta_{ab}})$ is the HD angular correlation from Eq.~\eqref{eq:HDcurve}, which includes the doubling of the autocorrelation via the ``pulsar'' term.
Note that fixing the injected angular correlation to Eq.~\eqref{eq:HDcurve} is equivalent to fixed injected values of each $c_{\ell}$, given by Eq.~\eqref{eq:LegendreCoefficientscale}, for all realizations.
The SGWB-induced timing residual for pulsar $a$ can then be expressed in the frequency domain as
\begin{equation}
  \tilde{R}^{\rm gw}(\hat{n}_a, f) = c(f) \: \sum_b M_{ab} \: w_b(f),
  \label{eq:Cholesky}
\end{equation}
where $c(f)$ is a frequency-dependent normalization chosen so that the two-point correlation of $\tilde{R}^{\rm gw}(\hat{n}_a, f)$ gives the SGWB power spectrum from Eq.~\eqref{eq:GWpowerlaw}, and $w_b(f)$ is a complex (two-parameter) Gaussian variable for pulsar $b$ at frequency $f$.
In practice, only a finite number of frequency components are modeled; the frequency components range from 0 to the Nyquist sampling frequency based on a 14-day cadence, in increments of $1/(10\:T_{{\rm obs}})$.
For each frequency component, the real and imaginary parts of $w_b(f)$ are independently sampled from a standard normal distribution so that the two-point correlation is given by
\begin{equation}
  \left< w_a^*(f) \: w_b (f') \right> = \frac{2}{T_{{\rm obs}}} \: \delta(f-f') \: \delta_{ab}.
  \label{eq:GPvector}
\end{equation}
The SGWB-induced timing residual $\tilde{R}^{\rm gw} (\hat{n}_a, f)$ is transformed back into the time domain, and interpolation is used to find the value of ${R}^{\rm gw} (\hat{n}_a, t_{ai})$ at observation time $t_{ai}$, which is then added to the $i^{th}$ TOA for pulsar $a$.

To summarize, four different pseudorandom number generator seeds are specified for each mock PTA dataset realization.
To perform realization analyses, we vary the set of four generator seeds while ensuring the four generator seeds are all different within a realization.

%%%%%%%%%%%%%%%%%%%%%%%%%%%%%%%%%%%%%%%%
\subsection{Realization study results}
\label{sec:seed-results}

Fig.~\ref{fig:SDseeds} provides a comparison of the multipole evidence from the realization analyses in this appendix against the single realization in Sec.~\ref{sec:results}.
We calculate the multipole evidence using the Savage-Dickey Bayes factor as discussed in Appendix~\ref{app:savage-dickey}.
We see in Fig.~\ref{fig:SDseeds} that the single realization in Sec.~\ref{sec:results} has either the same evidence or conservatively low evidence, relative to range of potential evidence from the realization analyses.

In Fig.~\ref{fig:CIseeds}, we plot the mean and 68\% CI of parameters $\log_{10} A_{\text{gw}}$, $c_2$, $c_3$, and $c_4$ for each individual realization.
We see that the average of the means tends toward the injected value with increasing $T_{\rm obs}$ and $N_p$.
We can also see the mean-value fluctuation scale decreases with increasing $T_{\rm obs}$ and $N_p$, and we observe that the scale of the mean fluctuation $1\sigma$ error bars (shown as black dashed lines) is the same scale as the 68\% CI (shown as error bars around each plotted point).
In other words, the scale associated with the spread of a detected parameter's posterior 68\% CI, which is realization independent, is the same scale as the fluctuation of the parameter's posterior mean, which is realization dependent.
Our result is consistent with an ergodic process for SGWB signal fluctuations and explains an additional observation we make with the quadrupole posterior distributions from the realization analyses: with all three datasets, the 68\% CI for the quadrupole posterior distribution contains the injected value of the quadrupole approximately 68\% of the time.

A final result to mention from the realization analyses is that the realization-averaged standard deviations of the marginalized 1D posterior distributions for multipoles that are not detected are all the same for a given dataset: approximately 0.021, 0.014, and 0.012 for 50 pulsars with 10 years, 50 pulsars with 20 years, and 100 pulsars with 20 years, respectively.
These values coincide with the standard deviations of the posterior distribution for $\log_{10} A_{\text{gw}}$ in each dataset, which means $\sigma_{\log_{10} A_{\text{gw}}}$ is related to an inherent harmonic analysis uncertainty for multipoles that are not detected.
The parameter $\log_{10} A_{\text{gw}}$ is detected in all analyses, so the realization-dependent fluctuations of $\sigma_{\log_{10} A_{\text{gw}}}$ are small; hence, we can accurately predict this inherent uncertainty with a single realization.
Since $c_{6, {\rm injected}} \approx 0.0116$ is less than $\sigma_{\log_{10} A_{\text{gw}}}$ for these analyses, we can see why multipoles $\ell \ge 6$ are not detected in these realization analyses.
The combination of the spread for nondetected multipoles being comparable to $\sigma_{\log_{10} A_{\text{gw}}}$, along with an ergodic process for the realization-dependent fluctuations, explains why we occasionally see large values for undetected multipoles, such as $\ell=7$ for the 50 pulsar analyses in Sec.~\ref{sec:results}.

\subsection{Distribution of quadrupole realization means}
\label{sec:seed-variance}
\begin{figure*}[t]
  \includegraphics[width=0.98\textwidth]{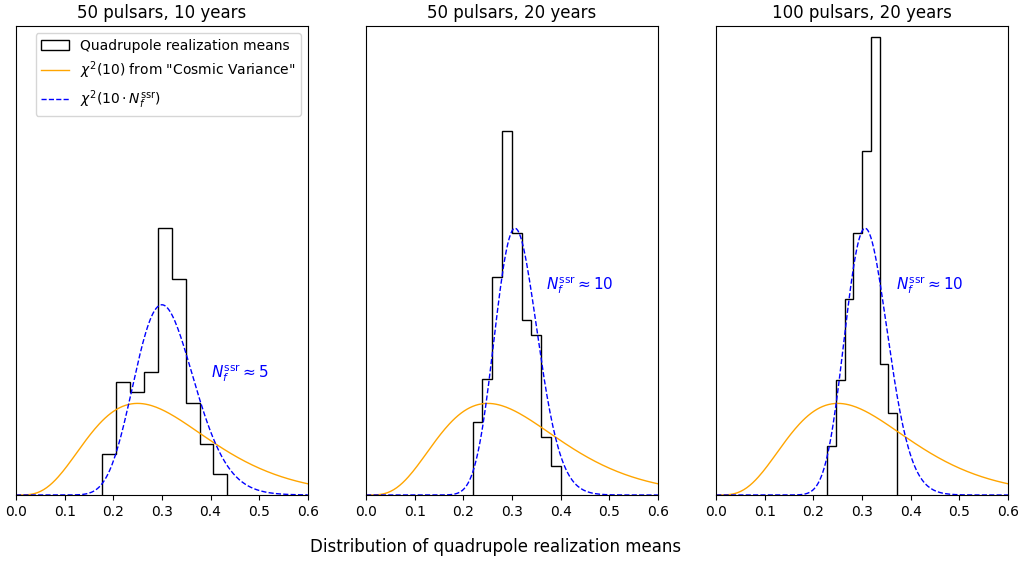}
  \caption{Comparison of distribution of quadrupole realization-dependent means to a $\chi^2$ distribution.
    Each of the three plots shows the results for a different realization study as indicated in the plot title.
    The less peaked $\chi^2$ distribution (orange curves in each plot) shows the ``cosmic variance'' for a single frequency bin~\cite{Roebber:2016jzl, Allen:2022dzg, Allen:2022ksj}.
    The ``cosmic variance'' degrees of freedom is 10 for the quadrupole $\chi^2$ distribution, which we explain in the main text.
    The more sharply peaked $\chi^2$ distribution (blue dashed curves) is obtained when the degrees of freedom scale linearly with the number of frequency bins in the strong-signal regime, $N_f^{\rm ssr}$.
    The signal-dominated frequency bins are determined by counting the number of frequency bins in our mock data where the signal-to-noise ratio is at least 10 (using the average values of the injected source and noise signals).
    Each $\chi^2$ distribution has been rescaled so that its mean value equals the value of the quadrupole coefficient from Eq.~\eqref{eq:LegendreCoefficientscale}.
  }
  \label{fig:Chisquared}
\end{figure*}
Previous work has shown the spatial correlation function has a minimum amount of sample variance (i.e., ``cosmic variance'') for a single frequency~\cite{Roebber:2016jzl} or an equal-time (i.e., zero-lag) autocorrelation function~\cite{Allen:2022dzg, Allen:2022ksj}.
According to this work, the (noise-free and infinite pulsar) angular power spectrum is a $\chi^2$ distribution with $2(2\ell+1)$ degrees of freedom for each multipole \cite{Roebber:2016jzl, Allen:2022dzg}, with standard deviation $c_{\ell} / \sqrt{2\ell+1}$, where $c_{\ell}$ is the value of the multipole coefficient from Eq.~\eqref{eq:LegendreCoefficientscale}.
In particular, the ``cosmic variance'' degrees of freedom for the quadrupole $\chi^2$ distribution is~10, corresponding to a quadrupole standard deviation between realizations of approximately 0.14 for a single frequency.

In Fig.~\ref{fig:Chisquared}, we plot a histogram of the quadrupole posterior mean values for the three different realization studies.
The orange curves in Fig.~\ref{fig:Chisquared} show a $\chi^2$ distribution with 10 degrees of freedom, which is the expected distribution from ``cosmic variance'' of the quadrupole.
Each $\chi^2$ distribution has been rescaled so that its mean value is equal to the injected value of the quadrupole given by Eq.~\eqref{eq:LegendreCoefficientscale}.
This comparison shows that our Bayesian harmonic analysis approach has a narrower distribution than expected from ``cosmic variance,'' even with our modeling of noise and a finite number of pulsars.

A possible explanation for this difference is that our analysis uses all available data when determining the posterior distribution of the covariance matrix, as opposed to using a single frequency bin or an equal-time estimator.
Since the sources that contribute to different frequencies of the SGWB are independent, each frequency bin provides an approximately independent sky map (see discussions in Sec. 4.3 of Ref.~\cite{Roebber:2016jzl}, Appendix~C.3 of Ref.~\cite{Allen:2022dzg}, and Appendix~B.2 of Ref.~\cite{Ali-Haimoud:2020ozu}).
Reference~\cite{Roebber:2016jzl} suggests that the degrees of freedom of the multipole $\chi^2$ distribution should scale with the number of frequency bins for a noise-free infinite pulsar dataset.

Since our analyses include intrinsic noise and use a finite number of pulsars, we expect only signal-dominated frequency bins to effectively contribute to these degrees of freedom. To determine the number of frequencies in the strong-signal regime, $N_f^{\rm ssr}$, we count the number of frequency bins where the signal-to-noise ratio is at least 10 (using the average values of the injected source and noise signals). For our mock data, $N_f^{\rm ssr} \approx 5$ for 10 years of observation, and $N_f^{\rm ssr} \approx 10$ for 20 years of observation.
Using a lower signal-to-noise ratio cutoff does not change these results appreciably.

The dashed blue curves in Fig.~\ref{fig:Chisquared} show a $\chi^2$ distribution with $10 \cdot N_f^{\rm ssr}$ degrees of freedom.
We can see that our realization-study results are much more consistent with this distribution than with the orange, ``cosmic variance,'' distribution. A similar result has been found in Ref.~\cite{Caliskan:2023cqm}, where it was observed that the ``cosmic variance'' is consistent with a single-frequency model of the data, but combining multiple frequencies reduces the total variance (e.g., see Fig.~8 of Ref.~\cite{Caliskan:2023cqm}).
We leave further investigations into this topic to future work.

%%%%%%%%%%%%%%%%%%%%%%%%%%%%%%%%%%%%%%%%%%%%%%%%%%%%%%%%%%%%%%%%%%%%%%%%%%%%%%%
\section{Savage-Dickey Multipole Evidence}
\label{app:savage-dickey}

We use the Savage-Dickey approach~\cite{10.2307/2958475} to calculate a Bayes factor for each multipole in our harmonic analyses, which is a measure of the evidence for the multipole in our model.
We calculate a Bayes factor by comparing the hypotheses $\mathcal{H}_{1,\ell}: c_{\ell} \neq 0$ against the hypothesis $\mathcal{H}_{2,\ell}: c_{\ell}=0$.
Our hypotheses are nested, $\mathcal{H}_{2,\ell} \subset \mathcal{H}_{1,\ell}$, because the prior for $c_{\ell}$ includes 0.
Moreover, since the prior for $c_{\ell}$ is uniform on its range, the prior probability distribution is $p(c_{\ell})=1$ for all prior values of $c_{\ell}$.
The Savage-Dickey Bayes factor for multipole $\ell$ is therefore given by
\begin{equation}
  \mathcal{B}_{\ell} = \frac{p(c_{\ell}\!=\!0)}{p(c_{\ell}\!=\!0\:|\:d, \mathcal{H}_{1,\ell})} = \frac{1}{p(c_{\ell}\!=\!0\:|\:d, \mathcal{H}_{1,\ell})},
  \label{eq:Savage-Dickey}
\end{equation}
where $d$ is the data and $p(c_{\ell}\!\!=\!\!0\:|\:d, \mathcal{H}_{1,\ell})$ is the marginalized 1D posterior distribution for parameter $c_{\ell}$ evaluated at $c_{\ell}=0$.

%%%%%%%%%%%%%%%%%%%%%%%%%%%%%%%%%%%%%%%%%%%%%%%%%%%%%%%%%%%%%%%%%%%%%%%%%%%%%%%
\section{Corner Plots}
\label{app:corner}

\begin{figure*}[t]
  \includegraphics[width=0.98\textwidth]{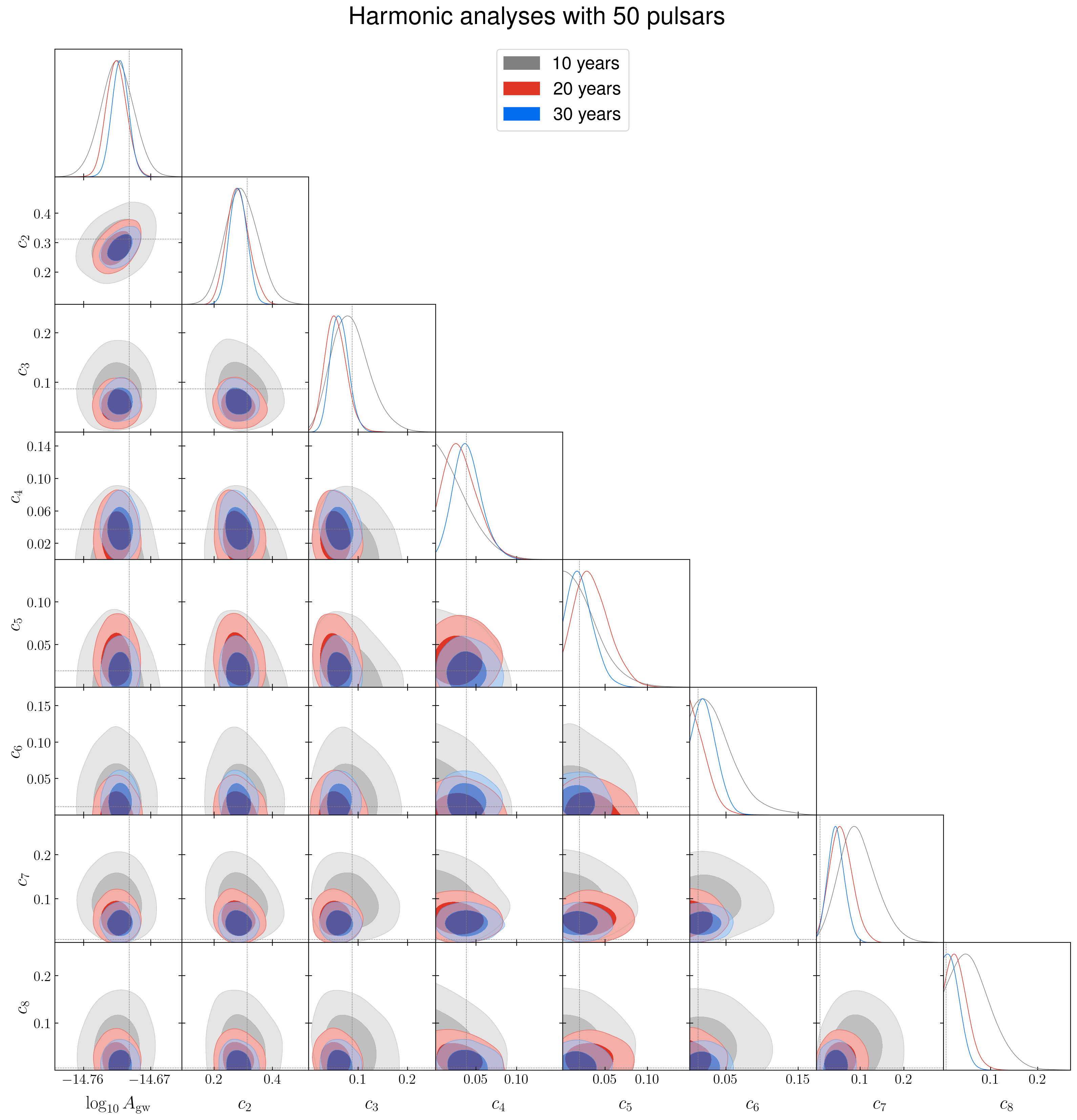}
  \caption{Corner plots for harmonic analyses of mock PTA datasets involving 50 pulsars.
    The 68\% and 95\% CIs of the marginalized posteriors of $\log_{10} A_{\text{gw}}$ and Legendre coefficients $c_\ell$ for total observation times of 10, 20, and 30 years are shown in gray, red, and blue, respectively.
    The injected values of the parameters are indicated by gray, dotted lines.
  }
  \label{fig:Corner1}
\end{figure*}

\begin{figure*}[t]
  \includegraphics[width=0.98\textwidth]{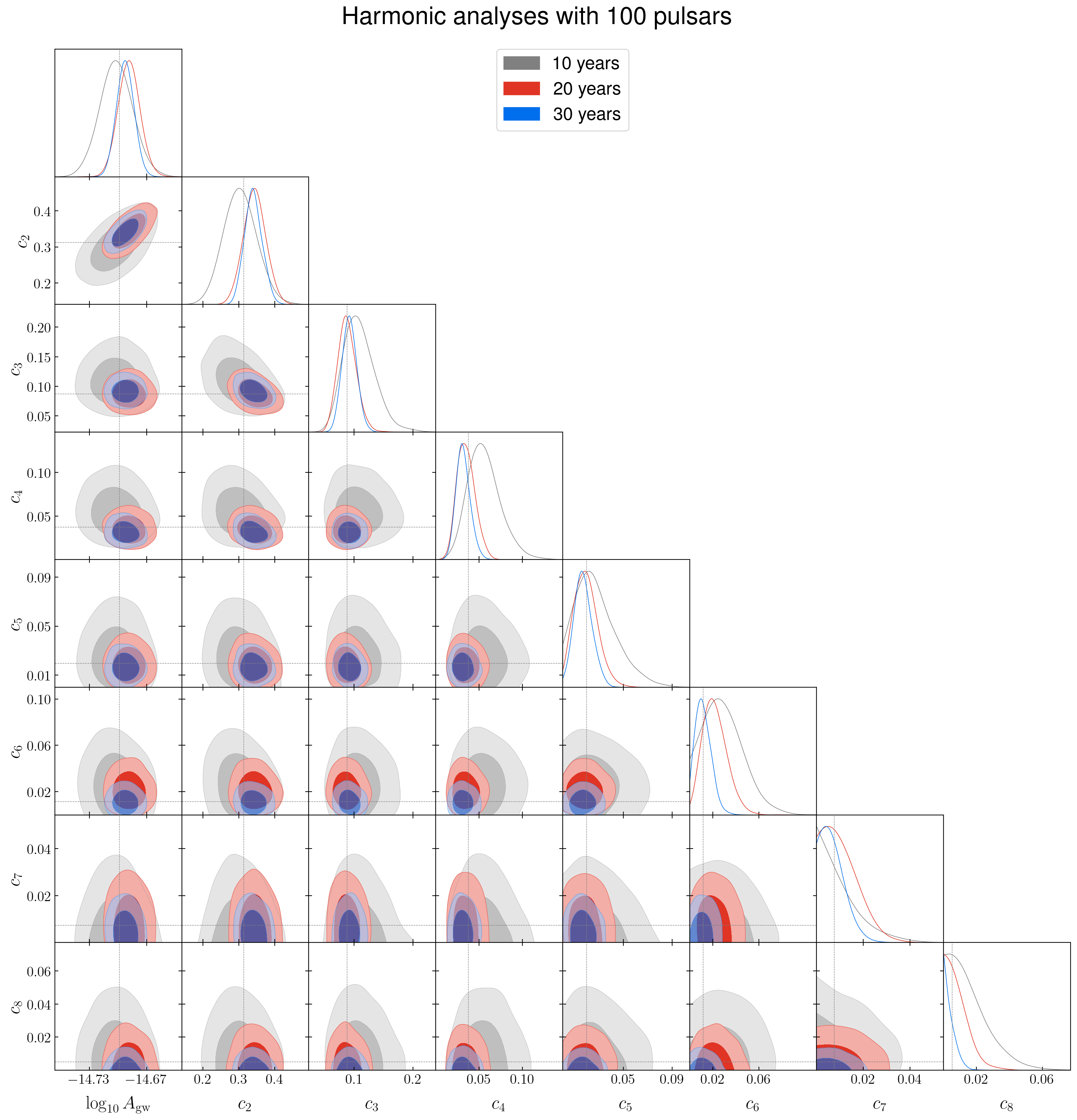}
  \caption{The same as Fig.~\ref{fig:Corner1}, except with 100 pulsars.}
  \label{fig:Corner2}
\end{figure*}

\begin{figure*}[t]
  \includegraphics[width=0.98\textwidth]{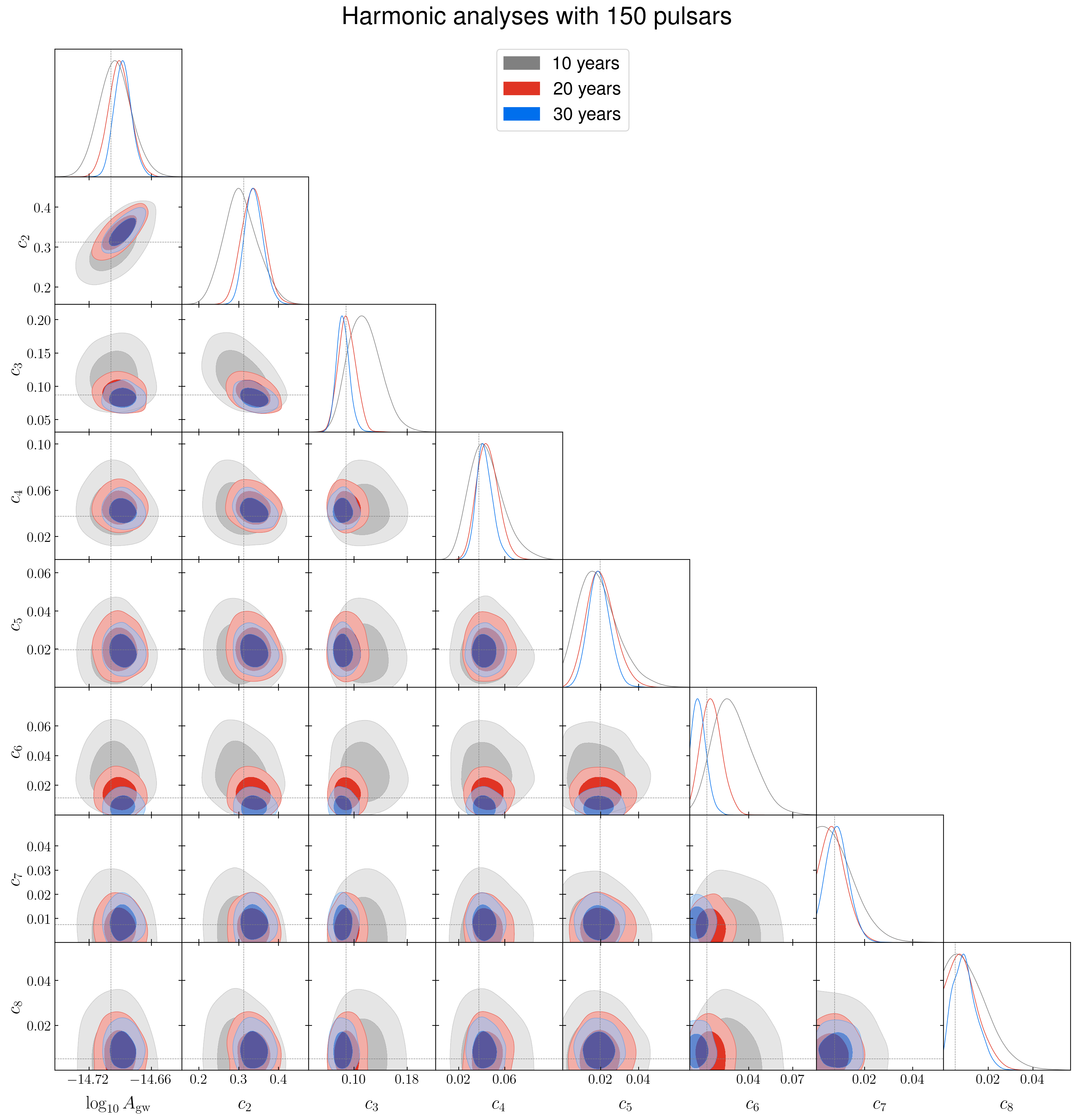}
  \caption{The same as Fig.~\ref{fig:Corner1}, except with 150 pulsars.}
  \label{fig:Corner3}
\end{figure*}

In this appendix we present the corner plots for the nine harmonic analyses discussed in Sec.~\ref{sec:results}.
Figs.~\ref{fig:Corner1}, \ref{fig:Corner2}, and~\ref{fig:Corner3} show the results of the harmonic analyses for 50, 100, and 150 pulsars, respectively.
In each figure we overlay the contours for 10, 20, and 30 years of observation time.

%%%%%%%%%%%%%%%%%%%%%%%%%%%%%%%%%%%%%%%%%%%%%%%%%%%%%%%%%%%%%%%%%%%%%%%%%%%%%%%%%%%%%%%%%%%%%%%%%%%%

\bibliography{harmonic}

%%%%%%%%%%%%%%%%%%%%%%%%%%%%%%%%%%%%%%%%%%%%%%%%%%%%%%%%%%%%%%%%%%%%%%%%%%%%%%%%%%%%%%%%%%%%%%%%%%%%

\end{document}